\documentclass{IEEE-Con-Sys-mag}
\def\BibTeX{{\rm B\kern-.05em{\sc i\kern-.025em b}\kern-.08em
    T\kern-.1667em\lower.7ex\hbox{E}\kern-.125emX}}
\usepackage{graphicx} 

\usepackage{graphics} 
\usepackage{epsfig} 
\usepackage{mathptmx} 
\usepackage{times} 

\usepackage{amsmath} 
\usepackage{amssymb}  
\usepackage{xcolor}
\usepackage{bm}
\usepackage{csquotes}
\usepackage{subcaption}
\usepackage{epstopdf}
\usepackage{color, soul}
\usepackage[compress]{cite}
\jvol{XX}
\jnum{XX}
\paper{8}
\jmonth{}
\jname{IEEE CONTROL SYSTEMS}
\pubyear{202X}

\begin{document}

\title{Recommender Systems as Control Systems\stitle{Ensuring fairness in dynamics of users and creators
}}

\author{{G}IULIA DE PASQUALE, SARAH DEAN, and PAOLO FRASCA}
\affil{}

\maketitle

\dois{}{}


\chapterinitial{R}ecommender systems (RS) are information filtering systems that provide the users of modern online platforms with personalized selections of items \cite{rs_basic}.  They generate profit by increasing user engagement with content, where the engagement can take the form of views, likes, re-sharing, or purchases, depending on the platform. Based on past observations from the users, RS generate predictive models of their behaviour from which they anticipate which items users are most likely to engage with. If we consider that content is typically made available on the platform by creators or sellers, in the case of e-commerce, the RS effectively act as intermediaries between creators and users. Nowadays, recommended content constitutes the overwhelming majority of the consumed content: its consumption can therefore influence users preferences and can exacerbate biases at both the user and content creator levels, affecting individuals as well as the broader population \cite{NP-JB-EE-GDP-SB-AH:23,mansoury2020,biases_review}. 

At the user level, by consistently promoting content that aligns with user preferences, RS can narrow the diversity of information users are exposed to, thereby nurturing the formation of echo chambers, namely environments in which a user encounters only beliefs or opinions aligned with their own, and the polarization of opinions.
Moreover, by targeting subpopulations of users who are more likely to engage with content, but are not fully representative of the overall population, such systems can discriminate against under-represented groups
~\cite{chowdary_Neurips25,Stoica_networkdiversity18,fangshuang2014diversified}. 

At the creator level, RS tend to promote creators who are more likely to drive engagement, thereby amplifying popularity bias and leading to rich-get-richer effects, which make it harder for new creators joining the platform to gain attention~\cite{sarah_arxiv24,ionescu_cc,ionescu_fairness_creators,ionescu2023luck}. Creators, in turn, shift toward formats and topics that are known to generate high engagement, e.g., because emotionally charged, sensational, or highly shareable~\cite{liang2025content}.

\begin{summary}
\summaryinitial{W}e propose a control‑theoretic interpretation of recommender systems and use this perspective to analyze how fairness interventions shape long‑term system behavior. Fairness concerns arise for both users and creators, ranging from opinion polarization and representation bias on the user side to popularity bias on the creator side. A central insight of our analysis is that fairness should not be viewed as a simple trade‑off against utility. When optimized over time, it can in fact be beneficial for overall system performance. Realizing these gains, however, requires a clear understanding of the underlying dynamics.
\end{summary}

\begin{figure}
    \centering
    \includegraphics[width=\linewidth]{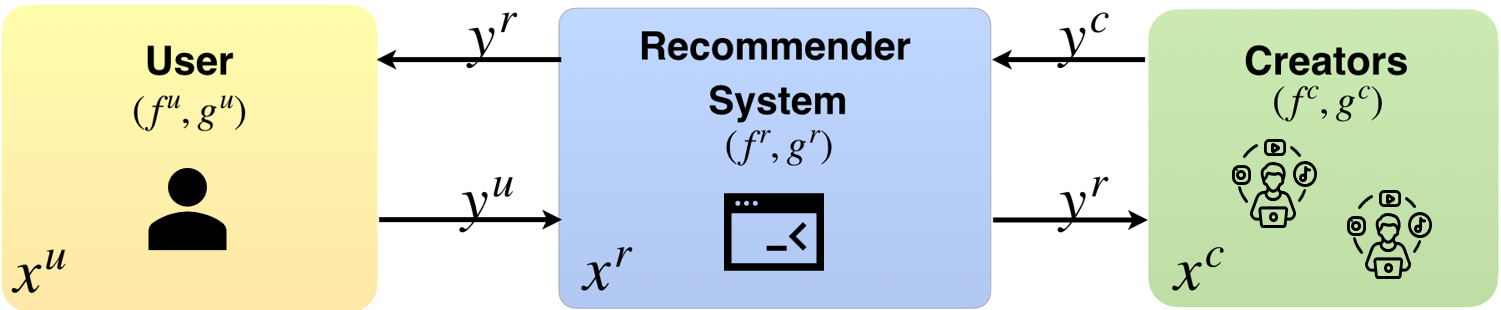}
    \caption{Formal interaction model for a content recommendation system, represented as a coupled dynamical system involving the recommender $(x^r, y^r, f^r, g^r)$, user $(x^u, y^u, f^u, g^u)$, and creators $(x^c, y^c, f^c, g^c)$. The variable $x$ denotes the internal state, the function $f$ describes the state evolution, the function $g$ represents the measurement equation, and the variable $y$ captures both inputs and outputs.}
    \label{fig:recsys}
\end{figure}

The underlying cause of these socially undesirable phenomena becomes clearer when adopting a systems-theoretic perspective and viewing the user–recommender–creator interaction as a coupled dynamical system, see Figure~\ref{fig:recsys}. Within this system, multiple \emph{positive feedback loops} emerge, driving the amplification of biases. In this framework, each stakeholder --namely, users,  RS, and creators-- is characterised by an internal state ($x$) and an output ($y$), both influenced by the recommended content $y^r$ as well as by the outputs of the other users. In particular, recommended content acts as an \emph{input} to users and creators, while their actions, such as purchases, dwell time, or produced content, constitute outputs that feed back into the RS through implicit feedback, thereby updating its predictive model. Formally, this interaction can be described through dynamical relations of the form $x^u_{t+1}=f^u(x^u_t,y^r_t)$ and $y^u_t=g^u(x^u_t,y^r_t)$, with analogous equations holding for the other stakeholders. Once this feedback loop is made explicit, the design of RS algorithms can naturally be interpreted as a control problem over a dynamical system. However, recommender algorithms are not traditionally designed in this way~\cite{SD-ED-MJ-LL:24}.

\section{Personalized Recommendation Models}

Modern RS typically rely on predictive models trained on large-scale historical interaction data. The core objective of these models is to estimate the likelihood that a user will engage with an item, such as clicking, likeing, purchasing, or watching it. Formally, the system learns a scoring function $s(u,i,c)$ that predicts the expected utility of recommending item $i$ to user $u$ in context $c$, where context may include factors such as time of the day, geographic location, device type or recent activity~\cite{rs_basic}. The model is trained using logged historical interactions across many users.  Because most real‑world platforms collect implicit feedback signals such as purchase history, dwell time, search patterns, rather than explicit ratings, the learning task is typically framed as click‑through‑rate prediction or utility estimation from observational data. Once a model has learned to assign a predicted relevance or utility score to each user–item pair, the basic recommendation strategy is to recommend the items with the highest predicted scores. In practice, this usually means generating a ranked list or a personalized feed, where items are ordered from most to least relevant according to the model’s estimates. This ranking‑based formulation underlies everything from e‑commerce product carousels to social‑media timelines: the system computes scores for a large candidate set, selects the top‑scoring subset, and presents them in descending order. 

\begin{figure}
    \centering
    \includegraphics[scale=0.3]{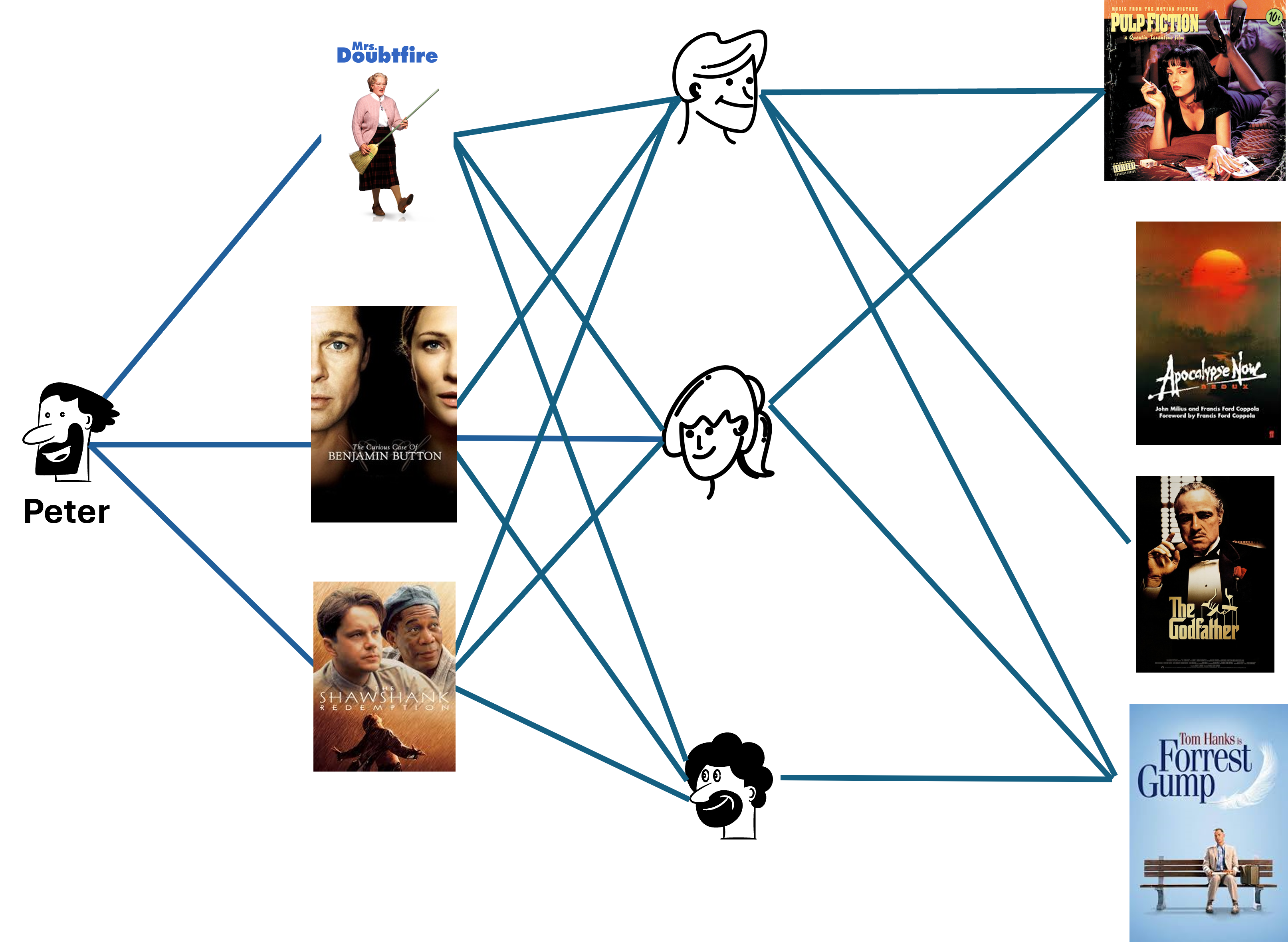}
    \caption{In the user-oriented neighborhood method, Peter likes the three movies on the left. To generate recommendations, the system identifies users with similar preferences (those who also liked these movies) and examines which additional movies they enjoyed. In this example, all three similar users liked \emph{Forrest Gump}, making it the top recommendation. Two of them also liked \emph{Pulp Fiction}, which becomes the next recommendation, and so on.}
    \label{fig:neighborhood}
\end{figure}

Historically, RS have been described through two broad modeling paradigms: \emph{similarity-based} methods and \emph{predictive models} \cite{rs_basic,collaborative_filtering}. Similarity‑based approaches dominated the early development of the field in the late 1990s and early 2000s. \emph{Content‑based} similarity methods construct a profile for each user and item to capture their meaningful characteristics. For example, the profile of a movie could include genre, actors, time of release etc. For users, profile might include demographic information. Such profiles are then used to generate recommendations by matching users to items with similar characteristics, thus maximizing the likelihood of the user to interact with the item. \emph{Collaborative filtering} similarity methods, instead, rely on patterns of historical user behaviour, assuming that users with similar interaction histories will prefer similar items. Classical collaborative filtering includes \emph{neighborhood‑based methods}, which predict a user’s preference for an item by examining ratings or interactions from similar users or similar items, and \emph{latent factor models}, which determines similarities based on content and users \emph{embeddings} into low-dimensional representations.  To be more clear, Figure~\ref{fig:neighborhood} replicates a similar example as in \cite{rs_basic} and illustrates an exemplification of neighbourhood method.  To predict the ratings of \emph{Forrest Gump} for the user, we would look at the movies nearest neighborhbors that Peter actually rated. The system then identifies like-minded users who can complement each other ratings. 
\begin{figure}
    \centering
    \includegraphics[scale=0.27]{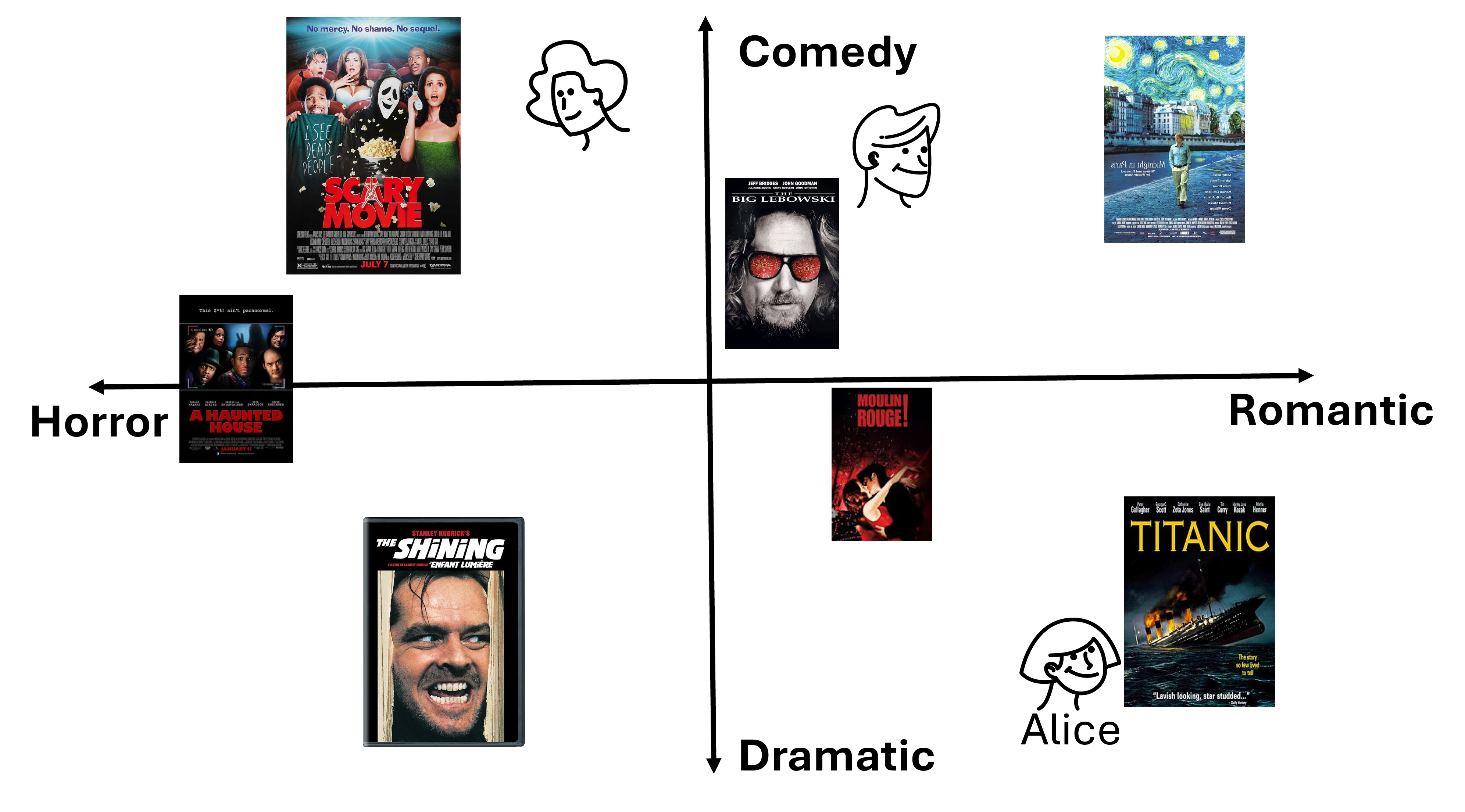}
    \caption{A simplified illustration of the latent factor approach represents both users and movies along two dimensions: horror vs. romance and comedy vs. drama.}
    \label{fig:latent}
\end{figure}

Beyond their algorithmic formulation, these early methods were shaped by the research communities that developed them. Much of the foundational work emerged from human–computer interaction (HCI) and information‑retrieval researchers, who emphasized not only predictive accuracy but also richer measures of user experience, such as perceived usefulness, trust, serendipity, transparency, and overall satisfaction. As a result, evaluation frameworks in this period often combined offline accuracy metrics with user studies and qualitative assessments. At the same time, the commercial potential of these systems became apparent as e‑commerce platforms grew. Companies such as Amazon and later Netflix rapidly adopted collaborative filtering techniques, accelerating the transition of RS from academic prototypes to large‑scale industrial infrastructure. These phenomena strongly influenced how similarity‑based methods evolved and set the stage for the predictive‑modeling era that followed.
A major shift, in fact, occurred during the Netflix Prize era (2006--2010), when RS moved from memory-based heuristics toward prediction-driven, model-based approaches. In this paradigm, the central objective became the accurate estimation of a scoring function \( s(u,i) \), i.e., predicting how a user \(u\) would rate or interact with an item \(i\). This marked a change in emphasis: rather than optimizing directly for holistic user experience or downstream engagement, much of the research effort focused on improving predictive accuracy on testing data, as reflected in the competition’s evaluation metric.

Among the approaches that emerged, latent factor models, and in particular matrix factorization, quickly became dominant. These models map users and items into a shared low-dimensional latent space inferred from historical interactions. Concretely, each user \(u\) is associated with a vector \(p_u \in \mathbb{R}^k\), and each item \(i\) with a vector \(q_i \in \mathbb{R}^k\), where \(k\) (typically between 20 and 100) represents the number of latent factors. The predicted score is then given by the inner product:
\[
s(u,i) = p_u^\top q_i.
\]
These latent dimensions can be interpreted as capturing underlying preference structure,  for example, a movie’s position on a comedy-drama spectrum or a user’s affinity for romance versus horror, though in practice they are learned implicitly from data rather than predefined.

This formulation provides an intuitive geometric interpretation: users and items that are ``close'' in the latent space (i.e., have high vector similarity) are more likely to be matched. For instance, in a simplified two-dimensional setting, by referring to Figure~\ref{fig:latent}, one might expect a user like Alice to align strongly with a movie such as \emph{Titanic}, while showing low affinity for a film like \emph{Scary Movie}.  

Among latent-factor methods, matrix factorization proved highly effective due to its scalability and strong predictive performance, which made it the cornerstone of many winning solutions in the Netflix Prize. Beyond that specific context, the core idea of learning dense latent representations remains highly influential in modern RS, including neural and deep learning--based approaches, where embeddings play a similar role.

Like most latent-factor approaches, matrix factorization inherits the statistical properties of the data on which it is trained. When interaction data exhibit systematic imbalances, such as popularity bias, uneven exposure, or under-representation of certain users or items, the learned latent representations encode these patterns. As a result, the model can propagate and even amplify such biases in its predictions, reinforcing disparities in visibility or recommendation quality across groups \cite{biases_review}. Understanding these dynamics is therefore essential when evaluating RS, especially in contexts where fairness and equitable exposure are important \cite{farnadi2020unifying,stoica2020seeding,karimi2018homophily,fish2019gaps,zhu2019group,tsang2019groupfairness}.

Modern recommender architectures extend the formulation of latent factor models to incorporate richer feature sets, yielding models of the form $s(u, i, c_u, c_i)$, where $c_u$ and $c_i$ represent user and item features such as demographics, device information, metadata, or content embeddings. Within this broader class of predictive models, a particularly influential development has been the emergence of two-tower (dual-encoder) architectures \cite{youtube, two_tower_model}, popularized by large-scale industrial systems such as YouTube's candidate-generation model. Nowadays, the two-tower approach represents the standard design, in which separate neural networks encode users and items into a shared embedding space, and recommendations are generated by computing the similarity, typically a dot product, between the two representations. This structure enables efficient large-scale retrieval via approximate nearest-neighbor search and naturally accommodates both collaborative signals (through learned embeddings) and content-based features (through feature-rich input layers). As a result, two-tower models have become a foundational component of modern retrieval systems, 
complementing more expressive ranking models that operate downstream. Recent research explores optimising for long-term outcomes, such as user satisfaction or retention, rather than short-term click signals, though data sparsity and delayed feedback make this difficult in practice\cite{implicit_feedback}.
Reinforcement learning approaches have also been proposed \cite{deep_learning_RS,rs_RL2}, treating the recommender as a policy and updating parameters via policy-gradient methods, but while promising, these methods remain largely experimental and are not yet widely deployed in production systems.
Despite these advances, fairness challenges remain and in fact several themes from earlier generations of recommender‑system research continue to shape modern practice. 

First, the conceptual tools developed around matrix factorization remain deeply relevant. Even though production systems now rely on large neural architectures, the geometric intuition of users and items occupying a shared latent space still provides a powerful way to reason about generalization, similarity structure, and the effects of data sparsity. Many industrial pipelines explicitly preserve this structure through embedding layers, factorized representations, or two‑tower architectures that behave as learned extensions of classical latent‑factor models.

Second, the increasing reliance on complex machine learning pipelines introduces new fairness challenges rooted in imperfect prediction. Models trained on observational logs can systematically underestimate the utility of items associated with minority users\cite{Liu_2023_groupsfairness}, niche interests\cite{liu2025_collaborative}, or historically under‑exposed creators \cite{liu2025identifying}. These misestimates propagate through ranking, leading to skewed exposure, reduced opportunity, and feedback loops that reinforce initial disparities. As a result, a substantial body of recent work focuses on diagnosing and mitigating bias arising from data imbalance and model misspecification.

Third, and more fundamentally, fairness concerns persist even in the hypothetical scenario where the scoring function is perfectly accurate. A perfect predictor of user utility does not imply a fair allocation of exposure. Ranking inherently concentrates attention at the top of the list, meaning that small score differences can produce large disparities in visibility. Position bias, limited user attention, and competition among items create structural inequalities that are not resolved by better prediction alone. Moreover, platform level objectives, such as engagement maximization, can interact with ranking mechanics to amplify popularity or suppress diversity. These issues arise from the allocation mechanism itself, not from model error, and therefore require interventions beyond improving predictive accuracy.

Together, these points highlight a central tension in modern recommender‑system design: while modeling sophistication continues to increase, many of the most persistent challenges: interpretability, bias, exposure inequality, and long‑term ecosystem health stem from structural properties of ranking and feedback rather than from the choice of model class.


\section{Fairness in Recommender Systems}

The literature on fairness in RS is extensive, encompassing a wide range of definitions, metrics, and mitigation strategies \cite{WY-MW-ZM-LY-MS:23, lanzettiCDC,misinformation_L4DC,NP-JB-EE-GDP-SB-AH:23,biases_review, rossi2021closed,HK-FY-SD:25}. Much of this diversity stems from the fact that RS operate in a closed-loop environment, where users, creators, and models continuously influence one another. These interactions generate multiple feedback loops that can amplify different forms of bias over time \cite{jiang2019degenerate,mansoury2020,NP-JB-EE-GDP-SB-AH:23}.
In the next sections, on the user side, we will restrict our focus on two illustrative mechanisms: an individual feedback loop \cite{perra2019modelling, rossi2021closed, SC-GDP-GB-FD:25, Sprenger_LCSS, LS-CAA-FD-GDP:26, damour2020fairness,heidari2019longterm,kleinberg2020classifiers,liu2020disparate,zhang2020fairdecisions} that intensifies historical bias \cite{historical_bias} and fuels opinion polarization, and a sampling feedback loop \cite{hashimoto2018fairness,zhang2020longterm,zhang2019group} that reinforces representation bias, which arises when the RS's development sample underrepresents some part of the population, and subsequently fails to
generalize well for a subset of the use population \cite{bias_ml}. On the creator side, we will examine a mechanism that amplifies popularity bias, generating a rich‑get‑richer dynamic. The underlying mechanism behind biases amplification lies in the fact that user engagement signals shape item ranking, which in turn determines future exposure and preferences. Likewise, creators whose content receives early engagement are more likely to be promoted, further entrenching popularity dynamics (incidentally, the rewards of early engagement also justify the phenomena of click farms or more generally using bots to simulate engagement~\cite{castaldo2024fake}). These distinct feedback processes lead to different forms of bias amplification. As a result, fairness metrics in the literature are often developed in an ad hoc manner, each tailored to the specific bias associated with a particular feedback loop\cite{bias_ml}.

Most RS are optimized primarily for engagement. From a control perspective, fairness can be interpreted as an additional objective that must be balanced against engagement. Rather than enforcing fairness as a hard constraint at every decision step, it is  more natural  to treat it as a soft constraint embedded in the objective function, allowing the system to optimize performance over a time horizon. This dynamic viewpoint highlights an important distinction between \emph{instantaneous fairness}, ensuring fairness at each individual recommendation step, and \emph{fairness over time}, where fairness guarantees are imposed over longer temporal windows. Static mitigation strategies, which enforce fairness at a single point in time, may therefore be ineffective in closed-loop systems because the underlying feedback mechanisms continue to operate and can reintroduce disparities resulting in suboptimal outcomes over a time horizon\cite{fairness_dynamic_ML}. 

This perspective naturally raises the question of how to trade off fairness and engagement over time\cite{Liu_delayedImpact_ICML18, hu2018shortterm,  HK-FY-SD:25}. Importantly, the relevant trade-offs depend on the timescale at which fairness is evaluated. While enforcing fairness at each step may significantly reduce short-term engagement, maintaining a balanced ecosystem of users and creators can improve platform health in the long run by promoting diversity, preventing creator churn, and sustaining user satisfaction. In this sense, fairness interventions may not only serve ethical objectives but also contribute to long-term system performance  \cite{AS-CGB-MP:23}.
In this tutorial paper, we adopt a control-theoretic viewpoint and focus on two notions of fairness that are particularly well suited to optimization for dynamical systems. Specifically, in the coming sections we  consider fairness from both the user perspective, where the goal is to mitigate the narrowing of information exposure to aligned content, and the creator perspective, where the objective is to prevent excessive concentration of attention due to popularity-driven feedback loops. These notions naturally lend themselves to formulations where fairness is enforced over time through dynamic objectives and control mechanisms, rather than through static constraints\cite{fairness_dynamic_ML}.

\begin{sidebar}{Biases in Recommender Systems}
    A concept closely related to fairness is the one of bias. RS are responsible for amplifying some biases which in turn can lead to fairness violations \cite{NP-JB-EE-GDP-SB-AH:23, WY-MW-ZM-LY-MS:23, mansoury2020}. Generally speaking, fairness is about how we want a RS to behave, while bias refers to errors or distortions that arise when a system learns patterns that do not accurately reflect the real world~\cite{WY-MW-ZM-LY-MS:23}. 
RS are trained and operate on observational user--item interaction data and are therefore exposed to multiple forms of bias. These biases arise at different stages of the recommendation pipeline and are further intensified by the closed feedback loop that links users, data, and models. This loop, in which the system determines what users see, users react to these exposures, and their reactions become new training data, see Figure~\ref{sfig2}, causes even small distortions that compound over time. Following the taxonomy in \cite{biases_review}, in what follows, we group the main biases into three categories: \emph{bias in data}, \emph{bias in model}, and \emph{bias in results}.

\sdbarfig{\includegraphics[width=19.0pc]{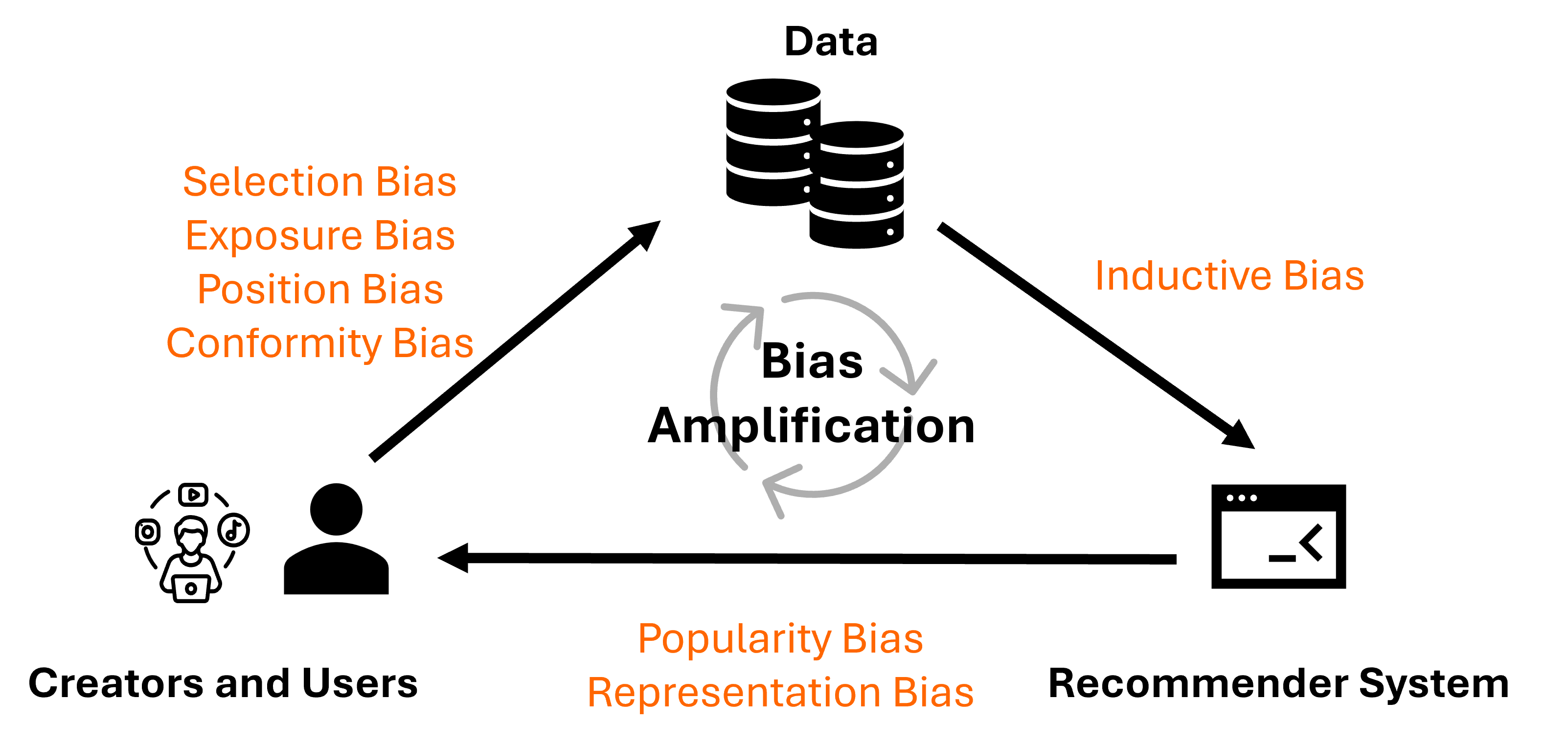}}{Feedback loop in the recommender system pipeline causing different biases amplifications.\label{sfig2}}

\subsection*{Bias in Data}
Biases in data originate during the collection of user interactions and reflect systematic deviations between the observed data distribution and the ideal distribution of all user--item pairs. Four prominent forms can be identified:
\begin{itemize}
    \item \textbf{Selection bias}: Users choose which items to rate or interact with, leading to missing-not-at-random data. Observed ratings or interactions are therefore not representative of all possible preferences.
    \item \textbf{Exposure bias}: Users can only interact with items they are exposed to. Unobserved interactions may reflect lack of exposure rather than lack of interest, making negative signals ambiguous.
    \item \textbf{Position bias}: Items at higher ranks receive disproportionate attention, regardless of their true relevance, producing skewed positive feedback for top-ranked positions.
    \item \textbf{Conformity bias}: User feedback can be influenced by social signals or public opinion, causing interaction labels to reflect conformity rather than genuine preference.
    \end{itemize}
Data-level biases distort the empirical distribution used for training, creating systematic errors that propagate into the learned model.

\subsection*{Bias in Model}
Beyond data issues, RS also incorporate \emph{inductive bias} through modeling choices. Such biases arise from architectural assumptions, regularization strategies, or feature representations introduced by system designers. While inductive bias is often necessary for generalization, it can interact with data imbalances in ways that favor certain items or users. For example, models may implicitly prioritize dense regions of the interaction space or overfit to popular items, reinforcing existing disparities.

\subsection*{Bias in Results}

During the serving stage, the model's outputs can exhibit two major forms of downstream distortion:

\begin{itemize}
    \item \textbf{Popularity bias}: Popular items receive more exposure, accumulate more interactions, and thus appear even more relevant to the model. This creates a rich-get-richer dynamic in which initial popularity advantages are amplified through repeated recommendation cycles.
    \item \textbf{Representation bias}: Certain user groups or item categories may systematically receive lower-quality recommendations or reduced visibility. These disparities often stem from upstream data imbalances but become more pronounced as the system repeatedly reinforces its own predictions.
\end{itemize}

\subsection*{Amplification Through Feedback Loops}

All seven biases are intensified by the feedback loop inherent in RS, as illustrated in Figure~\ref{sfig2}. The system's exposure decisions shape user behavior, user behavior shapes the training data, and the updated data reshapes future exposure. As a result, biases originating in data collection or model design do not remain static; they accumulate and propagate across iterations. Popular items become increasingly dominant, under-represented items remain obscure, and disadvantaged user groups receive progressively poorer recommendations. Understanding these amplification mechanisms is essential for developing fair and robust RS.
\end{sidebar}

\section{User-Side Feedback Loops}

Fairness issues in recommendation systems originate from dynamic, self-reinforcing, feedback-driven phenomena.
In this section, we begin their investigation by focusing on \emph{user-side} feedback loops. We illustrate their ability to degrade user representation on the platform and to exacerbate opinion polarization issues, among other consequences, see~\cite{NP-JB-EE-GDP-SB-AH:23} for additional examples of distinct downstream effects. Together with highlighting the issues, we present significant examples of countermeasures that suit our control-systems approach.

The first reason of concern is the {\em quality} of recommendations. Disparities in the quality of recommendations that users receive often originate from differences in user personalization models accuracy and uncertainties, which get amplified by the RS-user interaction \emph{positive} feedback loop. In fact, when personalization models are less accurate for certain user groups, these users receive less relevant recommendations, engage less and consequently generate less data, thus reinforcing the model's initial bias. 
In this regard, when there is high uncertainty about what a user is more likely to engage with, content diversity can both improve learning and provide fairer exposure.
A compelling perspective is to view fairness in RS design as a dynamic metric that accounts for the system operating in closed loop with users, where model predictions, ranking decisions, user behavior, and data collection continuously interact.
In this way, one can design mechanisms such as uncertainty-aware exploration, exposure allocation, and calibration, that prevent the amplification of disparities over time and promote equitable long-term outcomes.

Even in the presence of accurate predictions about user preferences, fairness concerns persist in terms of \emph{allocation of exposure}. Rankings that are purely driven by relevance can lead to narrow, self-reinforcing recommendation patterns, limiting the exposure to diverse content and thus contributing to the formation of filter bubbles and opinion polarization\cite{rossi2021closed,SC-GDP-GB-FD:25,LS-CAA-FD-GDP:26}. Also in this regard, as for the case of inaccurate prediction models, introducing diversity in the recommendation allocation can serve as a mitigation strategy for these undesired effects \cite{lanzettiCDC,rossi2021closed} by broadening the range of content that a user has access to. 

\subsection{User Representation}
RS interact with users through feedback loops that can shape not only individual behavior but also the composition of the user population on the platform \cite{representation_bias}. In particular, recommendation decisions may influence whether users remain active or leave the platform. When this interaction is repeated over time, the system dynamics can lead to unintended collective effects such as representation bias or population shifts across user groups.

A key mechanism behind these effects is \emph{sampling feedback loop} induced by platform participation \cite{NP-JB-EE-GDP-SB-AH:23}. Users remain active over the platform depending on the quality of recommendation they receive \cite{NP-JB-EE-GDP-SB-AH:23,hashimoto2018fairness,zhang2020longterm,zhang2019group}. Consequently, the set of observed users gradually becomes a biased sample of the underlying population. This sampling bias can be further reinforced by \emph{homophily}, the tendency of new users to resemble the current population. As a result, even if the platform initially reflects the population fairly, the long-term composition of users may drift toward a biased equilibrium.

To illustrate this phenomenon, let us consider the following stylized example taken from \cite{NP-JB-EE-GDP-SB-AH:23}. Suppose the population of users to be partitioned in two groups, \emph{Group~1} and \emph{Group~2}. The platform decides whether to recommend content to a user through a binary decision $d \in \{0,1\}$. When $d=1$, the user receives a recommendation and remains active on the platform. When $d=0$, the user receives no recommendation and leaves the platform. Initially, the population of active users is balanced, with approximately half of the users belonging to each group. Whenever a user leaves the platform, they are immediately replaced by a new user. To model homophily, the incoming user belongs to Group~1 and Group~2 with probabilities $p_1$ and $p_2$, respectively, where
\[
p_1 = \frac{n_{G_1}}{n}, \quad p_2 = 1-p_1, 
\]
where $n$ is the total number of active users. Thus, the more prevalent a group is on the platform, the more likely it is that new users belong to that group.

\begin{figure}
     \centering
     \includegraphics[width=\linewidth]{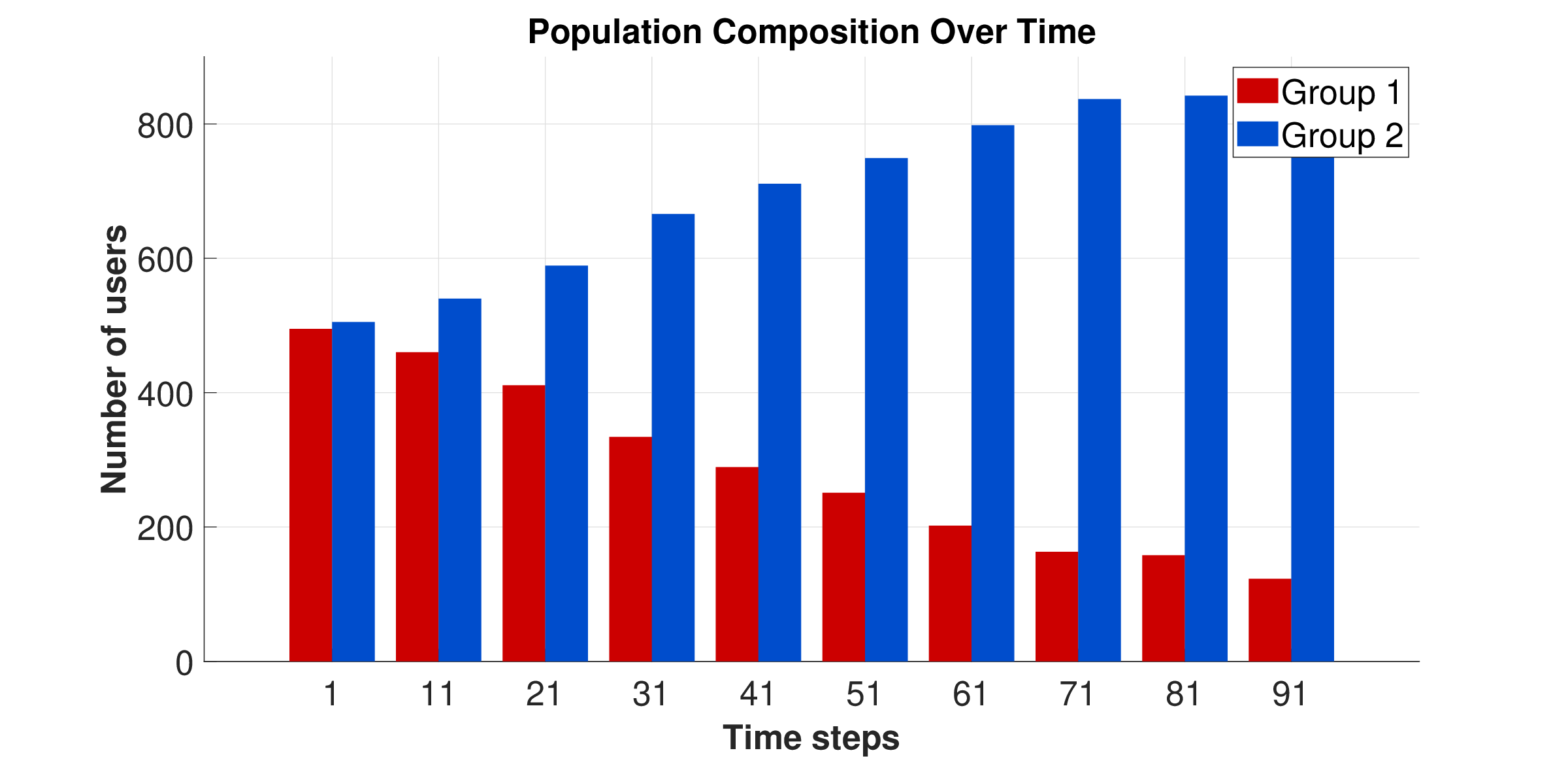}
     \caption{Representation bias propagated via the sampling feedback loop where an initally balanced population turns out in an over-representation of users from Group~$2$ as they are the ones with higher interest in the platform content.}
     \label{fig:representation_bias}
 \end{figure}

Over time, this feedback loop alters the population composition. As users who do not receive recommendations leave the platform, the active population becomes increasingly composed of users who are more likely to receive recommendations. Due to the homophily mechanism, the replacement process reinforces the current population distribution. As a result, even though the platform initially starts with nearly equal representation, the dynamics gradually reduce the number of one of the two groups of users, which, over time, may represent only a small fraction of the active population. The system then stabilizes around this biased composition, which corresponds to a stable equilibrium of the closed-loop dynamics. Figure~\ref{fig:representation_bias} represents a manifestation of this phenomenon, with a user population of $n=1000$ users, with $n_{\rm G1}=495$ users from Group~$1$ and $n_{\rm G2}=505$ users from Group~$2$, with Group~$2$ slightly more interested in the content platform, reflecting in a $50\%$ probability of receiving a recommendation, versus the $40\%$ for Group~$1$. As Figure~\ref{fig:representation_bias} shows, the self-reinforcing mechanism of untargeted users leaving the platform leads to a biased equilibrium point in the population distribution over time even when there was no representation bias in the initial population.

 Importantly, the bias does not necessarily originate from an initially unfair treatment of one group. Instead, it arises from the dynamics of the system itself: recommendations affect user retention, retention affects the observed population, and the observed population influences who joins the platform. Over time, these mechanisms can jointly lead to a persistent underrepresentation of certain groups \cite{NP-JB-EE-GDP-SB-AH:23,hashimoto2018fairness,zhang2020longterm,zhang2019group}.
 \begin{pullquote}
 Fairness interventions may not only serve ethical objectives but also contribute to long-term system performance. 
\end{pullquote}

\subsection{Opinion Evolution and Polarization}

From a user perspective, a fair RS promotes exposure to a broad range of information, including niche content. By diversifying the recommendations users receive, it can help break narrow information bubbles and patterns of repetitive content exposure, mitigate societal polarization, broaden users’ perspectives, and ultimately enhance the overall value of the recommendations~\cite{ben_rs}, thus mitigating the \emph{individual feedback loop} that forms between user and RS~\cite{NP-JB-EE-GDP-SB-AH:23}. To exemplify this notion of fairness, in the form of content diversity, let us consider the following scenario as in \cite{NP-JB-EE-GDP-SB-AH:23}. Consider a platform recommending a single type of content (e.g., political news). Each user $u$ has an opinion
\[
x_{u,t} \in [0,1]
\]
at time $t$, representing their level of interest or agreement with the topic. Values close to $1$ indicate strong interest, while values close to $0$ indicate strong disinterest.

\begin{figure}[htbp]
\centering
\begin{subfigure}{0.48\textwidth}
  \centering
  \includegraphics[width=\linewidth]{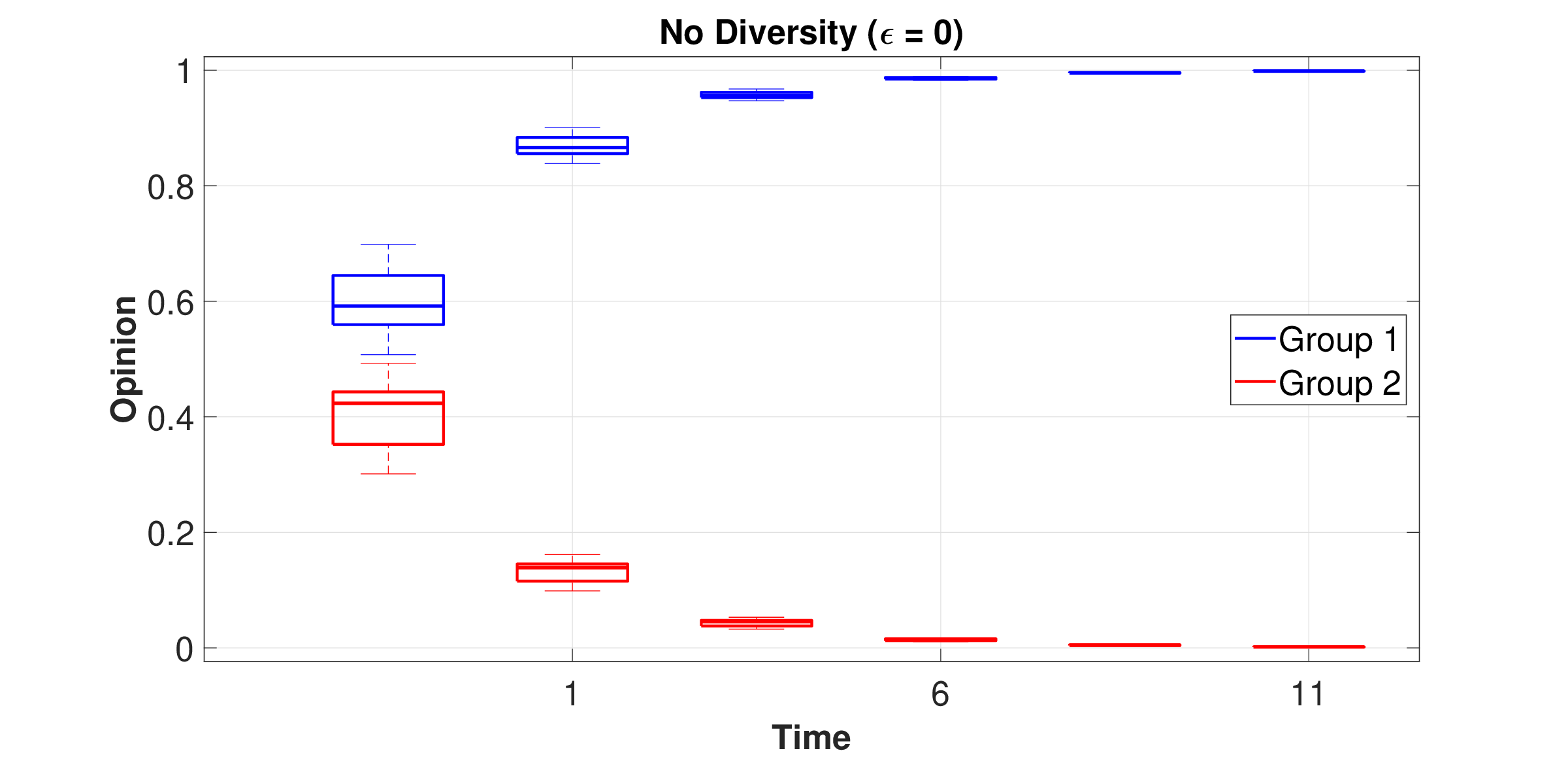}
  \caption{Opinion polarisation when no diversity is taken into account from the RS.}
  \label{fig:polarization}
\end{subfigure}%
\hfill
\begin{subfigure}{0.48\textwidth}
  \centering
  \includegraphics[width=\linewidth]{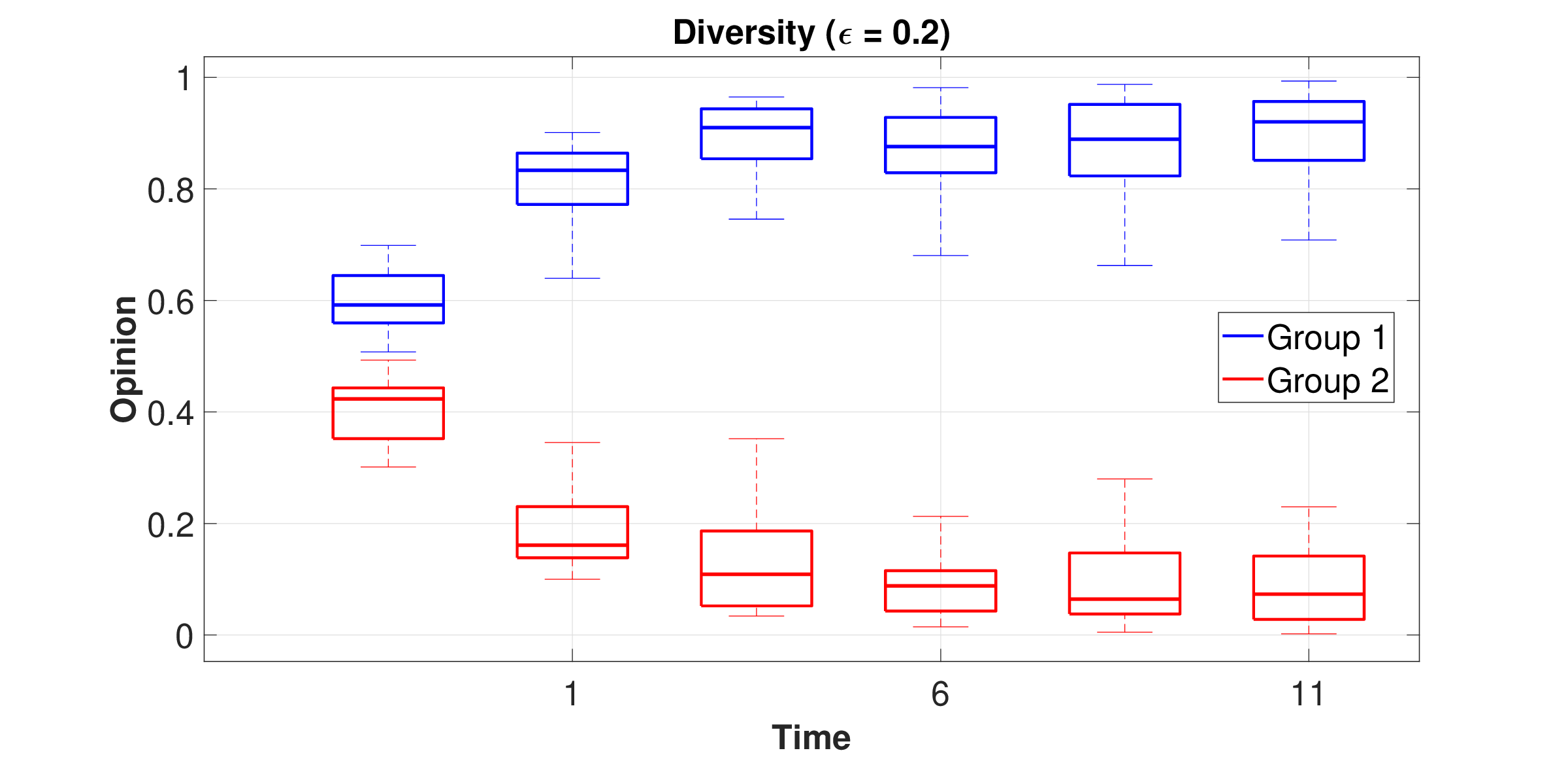}
  \caption{Content diversity mitigates opinion polarization.}
  \label{fig:diversity_polarization}
\end{subfigure}
\caption{Comparison of individual feedback-loop mechanisms: When no diversity is taken into account ( $\epsilon=0$), the self-reinforcing feedback loop causes opinion polarization towards extreme value (\ref{fig:polarization}). When diversity is taken into account, $\epsilon=0.2$, opinions fluctuate and the gap between groups is less prominent (\ref{fig:diversity_polarization}).}
\label{fig:test}
\end{figure}

At each time step, the RS selects content with stance $r_{u,t}~\in~[0,1]$ for user $u$. The user's opinion evolves according to the feedback rule
\begin{equation*}
x_{u,t+1} = (1-\alpha)x_{u,t} + \alpha r_{u,t},
\end{equation*}
where $\alpha \in (0,1)$ determines how strongly the recommended content influences the user.
Assume the RS aims to maximize engagement and therefore recommends content aligned with the user's current opinion:
\begin{equation*}
r_{u,t} =
\begin{cases}
1, & \text{if } x_{u,t} > 0.5, \\
0, & \text{otherwise.}
\end{cases}
\end{equation*}
This rule reflects a common engagement-driven strategy: users with stronger interest are shown more of the same type of content.
Suppose that we have two groups of users with slightly different initial opinion distributions:
\begin{itemize}
\item Group~1: $x_{u,0} \sim \mathcal{U}(0.5,0.7)$
\item Group~2: $x_{u,0} \sim \mathcal{U}(0.3,0.5)$,
\end{itemize}
with $\mathcal{U}(a,b)$, $a<b$ indicating the uniform distribution over the $(a,b)$ interval.
Even though the difference between the groups is small initially, the feedback loop amplifies it over time. Indeed,
for users with $x_{u,t} > 0.5$,
\begin{equation*}
x_{u,t+1} = (1-\alpha)x_{u,t} + \alpha,
\end{equation*}
which monotonically increases toward $1$, whereas, for users with $x_{u,t} < 0.5$,
\begin{equation*}
x_{u,t+1} = (1-\alpha)x_{u,t},
\end{equation*}
which monotonically decreases toward $0$.
As a result, the dynamics push users toward extreme opinions, see Figure~\ref{fig:polarization}. Over time, users with slightly higher initial interest converge toward $x=1$, while others converge toward $x=0$. The system therefore reaches a biased stable equilibrium where the population becomes polarized, reinforcing \emph{historical bias}.


To mitigate polarization, the platform can introduce content diversity by occasionally recommending content that differs from the user's current opinion. A simple approach is to randomize recommendations~\cite{rossi2021closed, lanzettiCDC}:
\begin{equation}\label{eq:rec_diversity}
r_{u,t} =
\begin{cases}
\text{aligned recommendation}, & \text{with probability } 1-\varepsilon, \\
\text{diverse content}, & \text{with probability } \varepsilon,
\end{cases}
\end{equation}
where diverse content is drawn from a broader distribution (e.g., $r_{u,t} \sim \mathcal{U}(0,1)$).
Under this policy, the opinion dynamics remain
\begin{equation*}
x_{u,t+1} = (1-\alpha)x_{u,t} + \alpha r_{u,t},
\end{equation*}
but now $r_{u,t}$ occasionally exposes users to different viewpoints.

This stochastic exposure prevents the deterministic drift toward extreme values. Instead of converging to $0$ or $1$, user opinions fluctuate around intermediate values determined by the balance between personalization and diversity. Consequently, opinion trajectories remain bounded away from extremes, and the difference between groups is less amplified over time, see Figure~\ref{fig:diversity_polarization}.

\begin{figure}
    \centering
    \includegraphics[width=\linewidth]{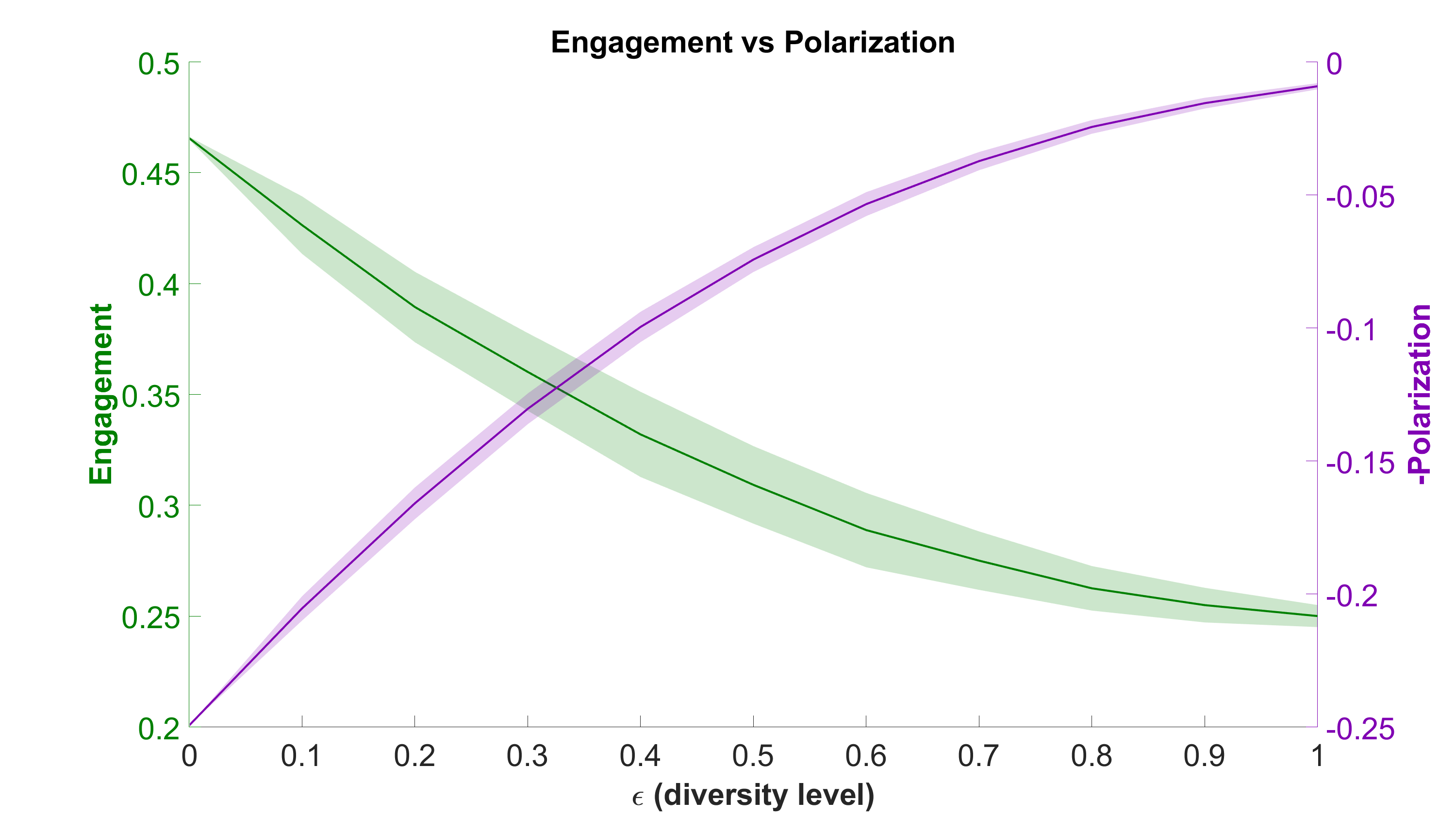}
    \caption{Trade-off between engagement and polarization as the content diversity parameter varies over 200 Monte Carlo simulations. Solid lines refer to the average, while the shaded area shows one standard deviation.}
    \label{fig:tradeoff}
\end{figure}

Thus, introducing diversity acts as a control mechanism that weakens the reinforcing feedback loop between recommendations and user opinions and needs to be balanced against engagement.
In fact, the expected engagement of user with opinion $x_u$, exposed to a recommendation $r_u$, can be assumed to be increasing with the alignment between state and recommendation. For simplicity, we may assume it to be proportional to $x_u \cdot r_u.$ It is therefore clear that, in this simplified setting where alignment is the sole driver of engagement, more diverse, less aligned recommendations are detrimental to engagement. Accordingly, Figure~\ref{fig:tradeoff} illustrates the relationship between engagement and opinion polarization in a simulated user population of $100$ individuals evolving over $30$ time steps under a RS with varying diversity level $\epsilon$. The results are obtained through a Monte Carlo simulation with $200$ independent trials, where each trial initializes heterogeneous user opinions, with intial opinion distributions split as before and evolves them according to stochastic recommendation and opinion update dynamics as in the previous dynamics \eqref{eq:rec_diversity}. The solid lines represent the mean engagement and negative polarization across trials for each value of $\epsilon$, while the shaded regions correspond to one standard deviation around the mean, capturing variability induced by random initialization and stochastic recommendation choices. The plot highlights how increasing $\epsilon$ systematically shifts the population from a highly engaged, but more polarized regime toward a less polarized but lower engagement regime. One needs to engineer RS to find the sweet spot in terms of the diversity parameter $\epsilon$ between engagement and polarization, which act as two competing forces. 

Of course, this randomization strategy is a crude way to address this trade-off. A more sophisticated approach, featuring item rankings, will be described in the next section. Other approaches have also been recently proposed, relying on more advanced control-theoretic tools (optimal control, online feedback optimization) to optimize costs that jointly evaluate both engagement and polarisation~\cite{mariano2026optimal,SC-GDP-GB-FD:25,LS-CAA-FD-GDP:26}.

Finally, we would like to note that in this discussion we have made the simplifying assumption that engagement is purely driven by the alignment between contents and preferences/opinions. In reality, the ``physics'' of engagement is richer, since in some contexts one might engage with content that they deem offensive, enraging, or otherwise opposite to their current opinion~\cite{brady2017emotion,rathje2021outgroup}.

\begin{pullquote}
    User preferences cannot always be understood as purely individual signals: They are often the outcome of collective social dynamics.
\end{pullquote}

\begin{sidebar}{Accounting for the Social Network to enhance Content Diversity}
\setcounter{sequation}{0}
\renewcommand{\thesequation}{S\arabic{sequation}}
\setcounter{stable}{0}
\renewcommand{\thestable}{S\arabic{stable}}
\setcounter{sfigure}{0}
\renewcommand{\thesfigure}{S\arabic{sfigure}}

\sdbarinitial{I}n much of the RS literature, users are modeled as independent decision-makers whose preferences evolve solely through their interactions with the platform \cite{rossi2021closed}. Under this assumption, recommendation algorithms treat each user in isolation, updating preference models based on implicit or explicit feedback. While this abstraction simplifies the design and analysis of algorithms~\cite{rossi2021closed}, it overlooks an important aspect of real-world information ecosystems: users are embedded in social networks through which they influence one another’s beliefs, attitudes, and preferences. Opinions are shaped not only by recommended content but also by discussions with friends, interactions on social media, and exposure to shared information within communities. As a result, user preferences cannot always be understood as purely individual signals; they are often the outcome of collective social dynamics \cite{kuhne_ICML25,musco_2018,musco_2020,chen_polarization,polarization_neurips}.

This observation has important implications when studying phenomena such as opinion polarization. Polarization is inherently a collective process arising from interactions among individuals within a network \cite{LS-CAA-FD-GDP:26,SC-GDP-GB-FD:25, Sprenger_LCSS}. Even if a RS were to diversify the content shown to each user individually, the resulting effect on collective opinions may still be limited if users primarily interact with like-minded peers. In such cases, social reinforcement mechanisms can amplify differences between communities, causing groups to drift further apart despite exposure to diverse content at the individual level. Consequently, interventions that operate solely at the level of individual recommendations may fail to address the structural mechanisms that sustain polarization \cite{user_creator}.
On the contrary, \cite{SC-GDP-GB-FD:25, LS-CAA-FD-GDP:26} have shown that taking into account the user's network can significantly improve
the performance trade-off between engagement and polarization (trade-off illustrated in  Figure~\ref{fig:tradeoff} in the main text). 

\sdbarfig{\includegraphics[width=19.0pc]{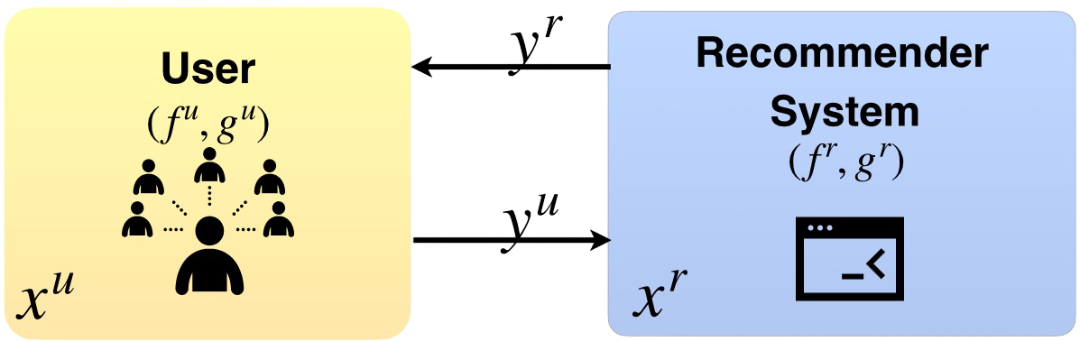}}{Shift in paradigm in RS design: Accounting for the user's social network plays a key role in mitigating undesired phenomena such as opinion polarization.\label{sfig1}}
These considerations suggest that mitigating polarization requires going beyond traditional diversification strategies. A more comprehensive approach must therefore incorporate the structure of the social network and the dynamics of opinion formation within it \cite{SC-GDP-GB-FD:25,LS-CAA-FD-GDP:26}, see Figure~\ref{sfig1}. By explicitly modeling how users influence each other, it becomes possible to study how recommendation policies affect not only individual engagement but also collective outcomes, such as consensus formation, fragmentation, or polarization.

From a systems perspective, this calls for a shift from user-centric optimization to network-aware RS \cite{SC-GDP-GB-FD:25,zhu2019group,SC-GDP-GB-FD:25,kuhne_ICML25,musco_2018,musco_2020,polarization_neurips,chen_polarization}. In such a framework, the RS becomes part of a larger dynamical system that includes both the recommendation algorithm and the social interactions among users. Understanding and controlling the emergent behavior of this coupled system is essential for designing recommendation policies that promote healthy information ecosystems while maintaining platform objectives such as engagement.
\end{sidebar}

\newpage

\section{Fairness as Content Diversity }
In the previous example, polarization was mitigated by diversifying recommendations uniformly at random.
This somewhat naive approach leads to trade-offs between engagement and polarization, as the diversity-inducing random selections may not be very relevant to users~\cite{ionescu2023luck}.
A more sophisticated approach instead formulates recommendation as an optimal ranking problem, incorporating diversity as an explicit constraint.
By doing so, the recommender can identify items that introduce a new viewpoint while still matching user preferences along other dimensions.
The result is a better trade-off between engagement and polarization than uniform randomization can achieve.

In this section, we elaborate on the perspective of diversity as a constraint on item rankings \cite{ionescu2023individual,pagan2021meritocratic}. 
This perspective also generalizes naturally from the user-side feedback loops discussed in the previous section to concerns on the content creator side. 
Indeed, fairness in RS is inherently multifaceted, reflecting the system's dual role as both an information filter for users and an exposure allocator for content creators. 

In the following, let $U$ be the set of users, $I$ the set of items, and 
$G = \{G_1, \dots, G_m\}$ a partition of items into groups. 
These groups can be defined based on  topic or ideological content, as in the polarization example, or based on other characteristics of the creator, such as protected attributes.
For each user $u \in U$, let $R_u$ denote the ranked list recommended to~$u$,
and let $R_u[k]$ be the item shown at rank $k$. 
Let $\pi(k)$ denote the position-based visibility (e.g., probability of observing rank $k$). 
Let $y_{u,i}$ be the observed user-response (e.g., click) to item $i$,
and let $s_{u,i}$ denote the latent relevance of item $i$ to user $u$.
We use $\mathbb{1}[\cdot]$ for the indicator function.
The notation is summarized in Table~\ref{table:notation}.
\begin{table}[t]
\centering
\caption{Notation.}
\begin{tabular}{ll}\label{table:notation}
\\
\hline
\textbf{Symbol} & \textbf{Meaning} \\
\hline
$U$ & Set of users \\
$I$ & Set of items \\
$G = \{G_1, \dots, G_m\}$ & Partition of items into protected groups \\
$u \in U$ & A user \\
$i \in I$ & An item \\

$R_u$ & Ranked list recommended to user $u$ \\
$R_u[k]\in I$ & Item shown to user $u$ at rank position $k$ \\
$\pi(k)$ & Probability user observes rank $k$ \\
$y_{u,i} \in \{0,1\}$ & Observed user response for item $i$ \\
$r_{u,i} \in \{0,1\}$ & Latent relevance of item $i$ to user $u$ \\
$\mathbb{1}[\cdot]$ & Indicator function \\
$\text{Exposure}(u,G_j)$ & Exposure of user $u$ to group $G_j$ \\
$\text{Exposure}(G_j)$ & Exposure allocated to group $G_j$ \\
$\text{Impact}(G_j)$ & Engagement with items in group $G_j$ \\
$\text{TPR}(G_j)$ & True positive rate for group $G_j$ \\
\hline
\end{tabular}
\end{table}

\subsection{Diverse Content Exposure}

As discussed in the section on user polarization, recommendations determine the information diet of users. 
When user opinions evolve as a result of the content they consume, relevance-based recommendations will lead to polarization. 
A natural remedy is to ensure that the ranked list of items presented to a user contains items from a diverse range of viewpoints. 
In this case, it is natural to consider the \emph{exposure} of a user to items from a particular group $j$:
\[
\mathrm{Exposure}(u,G_j)
=
\sum_{k=1}^{|R_u|}
\pi(k) \cdot \mathbb{1}[R_u[k] \in G_j].
\]
This expression sums the number of recommended items which are in group $j$, weighted by the position of those items in the ranked list.
To enforce a minimal level of exposure across all groups, the ranking problem could include a constraint of the form
\[
\mathrm{Exposure}(u,G_j)\geq \epsilon\quad \forall j
\]
The randomized policy from the polarization example would satisfy such a constraint in expectation, for two groups defined by their binary stance on a topic. 
However, by incorporating this constraint into an optimal ranking problem, one can achieve higher relevance while still meeting the diversity requirement, thereby mitigating the engagement/polarization trade-off.

\begin{sidebar}{Convex Relaxations for Optimal Ranking}
The goal is to maximize aggregate utility across all users:
\begin{equation*}
    \text{Utility} =  \sum_{u \in U} \sum_{k=1}^{n} r_{u,\, R_u[k]} \cdot \pi(k)
\end{equation*}
where we recall that $r_{u,i}$ denotes the item relevance to user~$u$, $\pi(k)$ is a position-based discount factor (e.g., discounted cumulative gain weights), $R_u[k]$ denotes the item placed at position $k$ in user $u$'s ranking, and $n$ is the length of $R_u$. 
Because this objective is linear and separable across users, the unconstrained optimum is simply to rank items in decreasing order of relevance for each user independently.

To promote fairness or diversity, we impose constraints.
While there is variety of possible types of constraints, a common one is to require that each group $G_j$ receives sufficient average exposure across users:
\begin{equation*}
    \max_{\{R_u\}_{u \in U}} \;\text{Utility} \quad \text{s.t.} \quad \text{Exposure}(G_j) \ge \epsilon_j \quad \forall\, j
\end{equation*}
where $\text{Exposure}(G_j)$, defined in the main text,
measures the average position-weighted visibility accrued by group $G_j$, and $\epsilon_j$ is a minimum exposure threshold.
To express the problem more compactly, we can represent the ranking for each user $u$ as a permutation matrix $\Sigma_u \in \{0,1\}^{n \times n}$, where $\Sigma_{u,ki} = 1$ if item $i$ is placed at position $k$. Let $\mathbf{r}_u \in \mathbb{R}^n$ be the vector of relevance scores for user $u$, $\boldsymbol{\pi} \in \mathbb{R}^n$ the vector of position weights, and $\mathbf{g}_j \in \{0,1\}^n$ an indicator vector with $g_{j,i} = 1$ iff $i \in G_j$. Then:
\begin{align*}
    \text{Utility}(u) &= \boldsymbol{\pi}^\top \Sigma_u\, \mathbf{r}_u,\quad 
    \text{Exposure}(G_j) = \frac{1}{|U|} \sum_{u \in U} \boldsymbol{\pi}^\top \Sigma_u\, \mathbf{g}_j
\end{align*}
Other diversity metrics discussed in this paper (user diversity, impact, opportunity) can be written as expressions with a very similar form but slightly different coefficients.

Optimizing directly over permutation matrices is combinatorially intractable. 
It is therefore common to relax $\Sigma_u$ to the convex hull of all permutation matrices---the Birkhoff polytope:
\begin{equation*}
    \Delta = \left\{\, \Sigma \in \mathbb{R}_{\ge 0}^{n \times n} \;\middle|\; \sum_{k=1}^n \Sigma_{ki} = 1 \;\;\forall\, i, \quad \sum_{i=1}^n \Sigma_{ki} = 1 \;\;\forall\, k \,\right\}
\end{equation*}
Any $\Sigma \in \Delta$ is a doubly stochastic matrix. 
Maximizing utility subject to exposure constraints over $\Sigma_u \in \Delta$ yields a linear program.
The LP solution $\Sigma_u^*$ is a doubly stochastic matrix, not necessarily a permutation matrix. 
However, by the Birkhoff--von Neumann theorem, any doubly stochastic matrix can be written as a convex combination of permutation matrices.
This decomposition can be computed efficiently and used to sample a discrete ranking. 
Under the resulting randomized policy, the utility and exposure constraints hold in expectation over the randomness of the sampled rankings. 
\end{sidebar}

\subsection{Creator Fairness}
Beyond allocating information to users, RS also play a resource allocation role for content creators by distributing exposure across items. 
The fairness of this allocation can affect both personal experience and broader social good \cite{WY-MW-ZM-LY-MS:23,ionescu2025_societal}. 
We now turn from the user side to the creator side and synthesize three major families of fairness notions for creators that have emerged in the recent literature \cite{WY-MW-ZM-LY-MS:23,review_fairnessRS_2}: equality of exposure, equality of impact, and equality of opportunity. 
These notions correspond to different normative commitments and different points of intervention in the recommendation process, and ultimately manifest as different operationalizations of content diversity.

\textbf{Exposure-based fairness} metrics aim to enforce diversity in what users are shown \cite{WY-MW-ZM-LY-MS:23}. 
By ensuring that different item groups receive comparable visibility, the system counteracts the concentration of attention on a narrow subset of popular items, allowing a broader range of products and niche items to surface more frequently. 
Formally, the exposure allocated to group $G_j$ is defined as:
\[
\mathrm{Exposure}(G_j)
=
\frac{1}{|U|}
\sum_{u \in U}
\sum_{k=1}^{|R_u|}
\pi(k) \cdot \mathbb{1}[R_u[k] \in G_j].
\]
Notice that this is the average of $\mathrm{Exposure}(u,G_j)$ across all users. 
This quantity sums the number of times content in group $j$ is recommended across all users, weighted by the ranked position.
A natural intervention, mirroring the user-side diversity constraint, is to require that each group receives a minimally guaranteed level of exposure: $\mathrm{Exposure}(G_j)\geq \epsilon$ for all $j$. 
A stricter notion is \emph{equality of exposure}, which requires $\mathrm{Exposure}(G_j)$
to be equal across all groups.
This means that, aggregated over the entire platform, each group is recommended equally often.

\textbf{Impact-based fairness} metrics enforce diversity in user interaction, rather than merely in what is shown. 
Since engagement (clicks, dwell time) is influenced by both exposure and user preferences, equalizing impact requires correcting for disadvantages that suppress engagement for certain groups. 
By ensuring equality of impact, the system creates opportunities for niche creators to accumulate feedback through a more balanced distribution of engagement across item groups. 
Thus, while exposure constraints diversify opportunities, impact constraints diversify realized outcomes. 
Formally, the realized engagement with items from group $G_j$ is:

\[
\mathrm{Impact}(G_j)
=
\frac{1}{|U|}
\sum_{u \in U}
\sum_{i \in G_j}
y_{u,i}.
\]
A common approximation is $y_{u,i} \approx \sum_{k=1}^{|R_u|}\pi(k)\, r_{u,i}\, \mathbb{1}[i=R_u[k]]$, which makes the connection to exposure explicit: impact then corresponds to exposure weighted by relevance $r_{u,i}$. Consequently, achieving equality of impact for a low-relevance group requires recommending its items more frequently to compensate.
As with exposure, one can either impose a minimum guarantee, $\mathrm{Impact}(G_j)\geq \epsilon$ for all $j$, or require full \emph{equality of impact}, meaning $\mathrm{Impact}(G_j)$ is equal across all groups.
This means that, aggregated over the entire platform, each group is engaged with equally often. 
Impact constraints tend to be stronger requirements than exposure constraints, since groups with lower relevance scores may need substantially higher recommendation rates to reach the desired engagement level.

Finally, \textbf{opportunity-based fairness} metrics 
measure how often a system recommends items from a given group conditioned on the relevance of those items (e.g., a user would click on it if shown). 
This notion aligns fairness with merit: if two items are equally relevant, they should have equal chances of being recommended. 
The practical effect is a guarantee of exposure among items the system deems relevant. 
Formally, the true positive rate for group $G_j$ is:
\[
\mathrm{TPR}(G_j)
=
\frac{
\sum_{u \in U} \sum_{k=1}^{|R_u|} \pi(k) \mathbb{1}[R_u[k] \in G_j]
}{
\sum_{u \in U} \sum_{i \in G_j} r_{u,i}
}.
\]
Similar to previous cases, we can consider both  a lower bound
($\mathrm{TPR}(G_j)\geq \epsilon$ for all $j$)
and \emph{equality of opportunity} ($\mathrm{TPR}(G_j)$ equal across groups).
The former ensures that each group receives exposure proportional to its underlying relevance, making it a less demanding requirement than the corresponding exposure or impact constraints. 
The latter equalizes the recommendation rate across all groups on the platform, conditional on relevance.

\begin{sidebar}{Relationship and Tensions Among Fairness Notions}
    Although equality of exposure, equality of impact, and equality of opportunity are often presented as distinct concepts, they are tightly coupled through the sequential structure of the recommendation pipeline, where the estimate of relevance shapes exposure, and exposure in turn shapes engagement. Because of this causal chain, improving fairness at one layer can positively affect the others: increasing opportunity fairness broadens the set of relevant items considered for recommendation, which tends to increase exposure diversity, while increasing exposure fairness boosts impact fairness by giving under‑represented items and groups more chances to accumulate interactions. At the same time, the notions can conflict and contradict each other. Specifically, equality of exposure does not guarantee equality of impact when user preferences differ, equality of opportunity may require re‑ranking that disrupts exposure parity, and equalizing impact may require intentionally unequal exposure. These complementarities and tensions reveal that the three fairness notions are best understood as different mechanisms for controlling diversity at different stages of the system, and that practical recommender design must navigate their trade‑offs through multi‑objective optimization rather than treating any single notion as sufficient on its own~\cite{review_fairnessRS_2}.
\end{sidebar}

\section{Creator-Side Feedback Loops}
From the creator's perspective, a platform's value lies in its ability to connect their content with the right audience. 
At the same time, platforms depend on creators to supply the diverse, high-quality content that keeps users engaged \cite{mansoury2020,rs_fairness_creators}.
It is therefore important to alleviate the
Matthew effect, namely the phenomenon where individuals or groups with initial advantages (wealth, status, or talent) gain more, while the disadvantaged fall further behind \cite{li2021user,ionescu_fairness_creators,ionescu_cc}. 
It may also be important to motivate these providers of niche items, and then improve the
diversity and creativity of items \cite{WY-MW-ZM-LY-MS:23, rs_item_diversity}. 
Fair RS handle these challenges by allocating more exposure (or impact, or opportunity) to long-tail items \cite{multi_stackeholder_fairness}.

We illustrate these ideas through a concrete example. 
Consider a platform with three types of users and three content creators (items).
Users differ both in their \emph{population size} (due either to raw differences in numbers or due to variation in the amount of traffic from that group on the site). User groups 1 and 2 are large groups (high traffic), while
group~3 is a smaller group. 
In particular, 
user group sizes are \(n_1 = 100\), \(n_2 = 100\), and \(n_3 = 10\).
Users also differ in their
preferences over content.
Creator~1 produces broadly appealing content for
user groups~1 and~2, Creator~2 produces niche content that is especially valuable to
user group~3 (and somewhat to group~2), and Creator~3 produces low-quality content that
is not preferred by any group.
The preferences are represented by a relevance matrix $r\in\mathbb R^{3\times 3}$ with entries \(r_{u,i}\), where \(u\in\{1,2,3\}\)
indexes user groups and \(i\in\{1,2,3\}\) indexes items: 
\[
r =
\begin{bmatrix}
0.9 & 0.1 & 0.0 \\
0.9 & 0.4 & 0.0 \\
0.2 & 0.9 & 0.1
\end{bmatrix}.
\]

Finally, we additionally consider the following feedback loop: creators stay on a platform only if they receive a sufficient level of exposure. 
Otherwise, they depart with some probability and cease creating content.
Such models have been studied in the literature~\cite{rs_matching,learning_rs}.
In this example, suppose the threshold is at $0.1$ and the retention probability is a sigmoid $p_i=[1 + \exp(-100(\text{Exposure}(i) - 0.10))]^{-1}$.
For simplicity, suppose that we recommend only one item to each user.
Then $|R_u|=1$
and the optimal recommendation problem can be simply written as 
\[
\max_{R_u\in\{1,2,3\}} \quad \sum_{u=1}^3 \sum_{i=1}^3 n_u \, r_{u,i} \, \mathbb{1}[ R_u=i]\:.
\]
Clearly, the optimal solution is to show content from Creator 1 to user groups 1 and 2 ($R_1=R_2=1$) and creator 2 to user group 3 ($R_3=1$) so that each group receives their maximal utility, which is $0.9$. rAs a result, Creator~1 receives an exposure of $\frac{200}{210}$, Creator~2 receives an exposure of $\frac{10}{210}$, and Creator~3 receives an exposure of $0$.
Due to the creator retention effects,
Creators~2 and~3 are likely to leave the platform and cease making content ($p_2\approx 0$, $p_3\approx0$). This reduces choices and leaves user group 3 with a much less preferred option.
One way to incorporate this issue into the utility definition is to define an expected future utility by replacing $r_{u,i}$ with the discounted $r_{u,i}\cdot p_i$.
Then it becomes clear that the maximal future utility of user group 3 drops from $0.9$ to $0.2$.

\begin{figure}[htbp]
    \centering
    \includegraphics[width=0.9\linewidth]{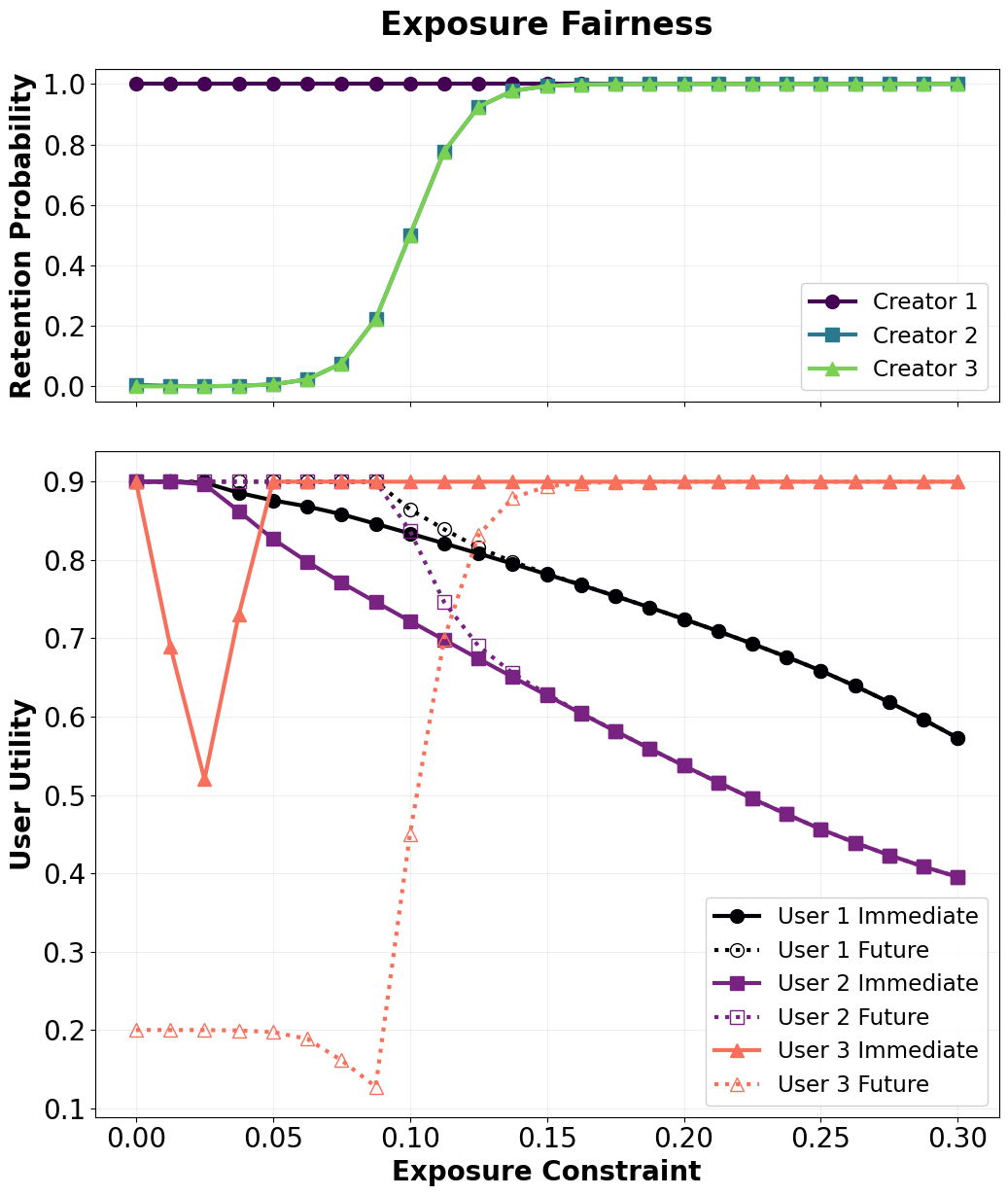}
    \caption{
    Outcomes for users and creators under recommendation with varying levels of exposure constraint. Top: as lower bound $\epsilon$ increases past the $10\%$ threshold, Creators 2 and 3 become likely to remain. Bottom: increasing $\epsilon$ results in lower immediate utility. Once the $10\%$ threshold is passed, the long term utility for user group 3 drastically improved, at modest cost to user groups 1 and 2.
    }
    \label{fig:group_exposure}
\end{figure}

A natural intervention is to require that each item receives at least a minimum
level of \emph{total exposure}. 
First, let us consider fractional recommendations within each of the three user groups, where $\Sigma_{u,i}$ is the fraction of user group $u$ recommended item $i$ (hence $\sum_{i=1}^3 \Sigma_{u,i}=1$).
Then formally, letting \(N = \sum_{u=1}^3 n_u\), we can impose for some \(\epsilon > 0\):
\[ 
\frac{1}{N}\sum_{u=1}^3 n_u \, \Sigma_{u,i}\;\ge\; \epsilon 
\quad \forall i.
\]
This constraint ensures that all creators receive a minimal level of exposure, including Creator 2.
However, it also forces exposure of Creator 3, which is low-value for all users.
To satisfy the constraint, the system must nevertheless recommend Creator 3, leading to a drop in overall immediate utility.
As shown in Figure~\ref{fig:group_exposure}, for values of $\epsilon$ above the $0.1$ exposure threshold, this immediate utility loss is the price for a much improved future utility for user group 3.

\begin{figure}[htbp]
    \centering
    \includegraphics[width=0.9\linewidth]{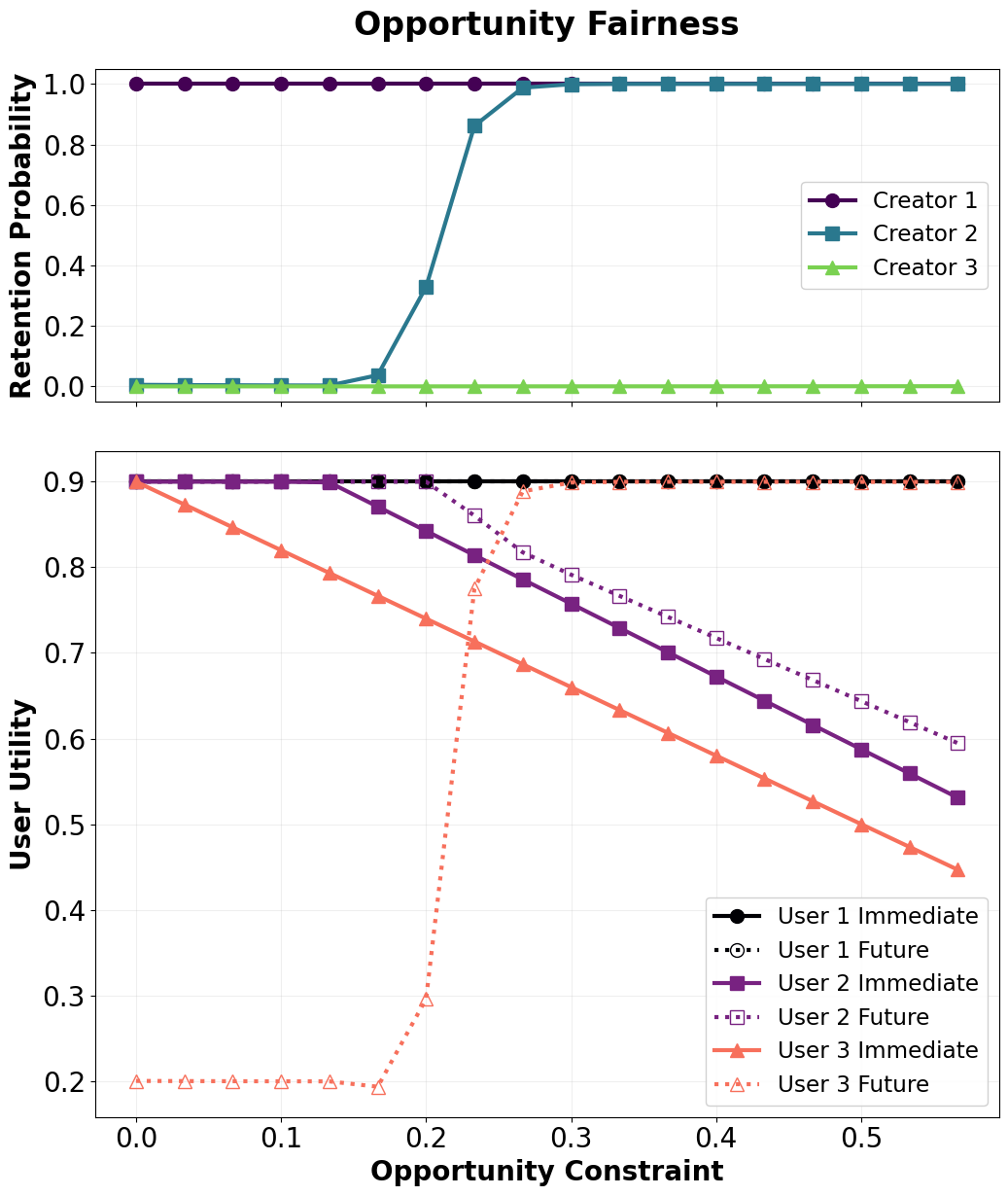}
    \caption{
    Outcomes for users and creators under recommendation with varying levels of opportunity constraint. Top: as lower bound $\epsilon$ increases, Creator 2 becomes likely to remain while Creator 3 never receives sufficient exposure. Bottom: increasing $\epsilon$ results in lower immediate utility, but higher long term utility for user group 3, at a more modest cost to user group 2.
    }
    \label{fig:group_opportunity}
\end{figure}

An alternative is to instead give creators a minimum level of \emph{opportunity}. 
Formally:
\[
\frac{1}{\sum_{u=1}^3 n_u r_{u,i}}\sum_{u=1}^3 n_u \, \Sigma_{u,i} \;\ge\; \epsilon 
\quad \forall i,
\]
for some \(\epsilon \in (0,1]\).
This constraint functions similar to an exposure constraint where the exposure level is altered for each group depending on their underlying relevance.
As a result, for modest values of $\epsilon$
the constraint enforces sufficient exposure for Creator 2
because they provide
high value to Group~3, without artificially promoting Creator 3. The resulting recommendations preserves useful diversity without
forcing low-quality content, leading to better long-term outcomes for user satisfaction, see Figure~\ref{fig:group_opportunity}.

In this example,
the opportunity-based intervention appears to better address the undesirable feedback loop wherein valued niche creators leave platforms.
However, depending on other factors at play, exposure constraints may also be important to
preserve low-demand creators (such
as Creator~3).
Considering the polarization example above, depicted in Figure~\ref{fig:polarization}, suppose that some viewpoint is only represented by Creator~3. Then they are necessary for mitigating polarization, even if relevance is lower.

More broadly, other feedback loops may exist between creator exposure and quality: 
those creators who receive sufficient support can grow their skills and produce more valuable content down the line with higher relevance
\cite{HK-FY-SD:25}.
Thus there are not always clearly better or worse ways to enforce content diversity. 
Even under the same system goals, the best fairness definition depends on the individual dynamics of user and creator behaviour~\cite{SD-ED-MJ-LL:24}. 
\begin{pullquote}
   Even under the same system goals, the best fairness definition depends on the individual dynamics of user and creator behaviour.
\end{pullquote}
Design choices about ranking constraints are essentially
control decisions which must be designed carefully with the underlying system dynamics and goals in mind. 

\section{Fairness over Time}

So far we have established that fairness, in the form of diversity constraints on recommended items,
can mitigate undesirable feedback loops.
Many of these concerns operate over 
the course of weeks or months as users opinions shift or creators decide to stop producing content. 
And we have seen in previous sections how the long term outcomes of fair recommendation policies can outperform those of myopically relevance maximizing ones\cite{fairness_dynamic_ML}.
However, recommendation algorithms need to operate on the much shorter time scale of minutes to serve user content.
In this section, we consider as a first-order factor the temporal aspect of fair recommendation.
First, we address the difference in time-scales between feedback loops relating to overall user satisfaction or platform health and recommendation decisions.
Then, we turn to the longest term outcomes which drive platform success and user satisfaction.


\subsection{Fast Time-scale Controllers}
On most platforms, recommendations to users are delivered on the order of minutes or seconds. 
A natural formulation of the fairness constraints discussed above requires that the total exposure (or impact or opportunity) delivered to each group
meets a prescribed target over a deployment horizon of $T$ recommendation steps.
At any individual step, satisfying a fairness constraint requires including lower-utility items into the ranking to achieve group-level fairness targets. 
This comes at a direct cost to relevance. 
Over the longer horizon $T$, the population of users visiting a platform changes, or the mood of users over time changes, and the relevance of items to those users varies accordingly. 
For example, a news platform might see heavier traffic from sports readers on weekends or political readers during election periods. 
These shifts create natural opportunities to satisfy exposure targets without necessarily paying a utility cost: if item group $1$ is highly relevant to Monday's users, it can accumulate exposure then, relaxing the constraint on Tuesday. Fairness over time is therefore less strict than fairness at every instant.
The constraint needs to be met only cumulatively, by the end of the horizon.

To view this as a control problem,
define the state $s_t$ as the quantity which tracks progress towards a fairness goal.
For example, the cumulative exposure delivered to each of $m$ groups up to time $t$ means that $s_t\in\mathbb{R}^m_+$.
At each step, the ranking controller observes the current context (item relevance scores based on active users), the state $s_{t-1}$, and must select a ranking that advances toward a target $\epsilon \in \mathbb{R}^m$ by the final step $T$.
One idea for achieving this goal is to adjust item relevance scores before sorting so that items belonging to groups that are not on track to reach their targets receive a boost.
A natural choice is to make this boost proportional to the tracking error $\tfrac{t}{T}\epsilon - s_{t-1}$, which measures how far each group is behind its linear interpolation target at time $t$.
Concretely, the adjusted score for user $u$ and item $i$ at time $t$ is
\[
    \tilde{r}_{u,i,t} = r_{u,i} + \gamma \sum_{j=1}^m \mathbb{1}[i\in G_j] \max \left(0, \tfrac{t-1}{T}\epsilon_j - [s_{t-1}]_j\right),
\]
where $\gamma$ is a gain parameter controlling sensitivity to tracking errors, and only tracking errors in one direction are of concern due to the direction of the inequality constraint.
Items are then simply ranked by $\tilde{r}_{u,i,t}$.
This approach, also known as proportional (P) control, was originally proposed as a heuristic for fairness in ranking~\cite{morik2020controlling}.
Its simplicity makes it practical for use under latency requirements.

Later work showed that
the P-controller is not merely a heuristic
\cite{sarah_arxiv24}.
In fact, it can be derived as a special case of online optimization
applied to the exposure-constrained optimal ranking problem, which maximizes utility (measured by relevance) subject to exposure constraints.
The Lagrangian form of this optimization problem is
\begin{align*}
    \max_{R_{u,t}} \min_{\lambda\geq 0}  &\frac{1}{T}\sum_{t=1}^T\sum_{u \in U} \sum_{k=1}^{n} r_{u,\, R_{u,t}[k]} \cdot \pi(k)\\ &-\sum_{j=1}^m \lambda_j\left(\frac{\epsilon_j}{T} -\frac{1}{T}\sum_{t=1}^T\mathrm{Exposure}_t(G_j) \right)
\end{align*}
The online optimization approach
alternates between max and min: selecting a ranking $R_{u,t}$ at each $t$ and then updating the Lagrange multiplier $\lambda$~\cite{agrawal2014fast}.
During the first step, the optimization reduces to:
\begin{align*}
    \max_{R_{u,t}}   &\sum_{u \in U}\sum_{k=1}^n \pi(k) \left(r_{u,\, R_{u,t}[k]} +\sum_{j=1}^m \frac{\lambda_j}{|U|}
 \mathbb{1}[R_{u_t}[k] \in G_j]. \right)
\end{align*}
which is solved by ranking the scores 
$r_{u,\, i} +\sum_{j=1}^m \frac{\lambda_j}{|U|}
 \mathbb{1}[i \in G_j]$.
Thus we see that the score boost depends on the multiplier $\lambda$.
In fact, if $\lambda$ is updated in the second step using gradient descent with step size $\gamma$,
Thus, the P-controller is exactly recovered.
This connection provides both a principled justification for the P-controller and a clear path to extensions.
PID-control could be interpreted as online gradient descent with adaptive step size and momentum \cite{hu2017control}.
Predictable temporal trends in user interests could be better exploited by replacing the linearly interpolated exposure target with forecasts of future item demand.
More broadly, viewing fairness in ranking through the lens of control theory opens a rich design space that is largely unexplored. 
\subsection{Long-term Outcomes}




 Fairness interventions cannot be evaluated solely in a static or one-shot manner \cite{fairness_dynamic_ML}; rather, their long-term impact emerges through feedback loops and equilibrium behavior \cite{SD-ED-MJ-LL:24,NP-JB-EE-GDP-SB-AH:23,rossi2021closed}.
A growing body of work shows that, in the absence of intervention, these feedback loops lead to self-reinforcing dynamics, such as popularity bias or “rich-get-richer” effects, which can lead to highly unequal steady states \cite{jiang2019degenerate,castaldo2024fake, rs_matching,learning_rs}. In such settings, even interventions that appear fair in the short term may fail to prevent, or may even exacerbate, disparities over time if they do not explicitly account for system dynamics \cite{damour2020fairness,heidari2019longterm,hashimoto2018fairness}.
In the prior section, we focused on satisfying fairness targets more flexibly through a ranking rule operating on fast timescales, translating between targets that matter on the scale of days or weeks. However, long-term outcomes unfold over months or longer, and this raises a deeper question: how should we set these targets to begin with?

When fairness is treated as a long-term control objective, a different picture emerges \cite{Liu_delayedImpact_ICML18, hu2018shortterm, AS-CGB-MP:23, HK-FY-SD:25}. Recent frameworks model fairness-constrained decision making as a sequential or dynamical optimization problem, where policies are optimized not only for immediate utility but also for their impact on future system states \cite{HK-FY-SD:25,kuhne_ICML25,rs_matching}.
By referring to the \emph{User Representation} example depicted in Figure~\ref{fig:representation_bias}, consider a RS that repeatedly selects which content to show to users from two groups, Group~$1$ and Group~$2$, where recommendations influence not only immediate engagement, but also future user activity. Let the system state $v_t \in [0,1]^2$ capture the fraction of active (or highly engaged) users in each of the two groups at time $t$. The evolution of the system depends on the recommendation policy~$\beta$, which determines the exposure allocated to each group.

A myopic recommender, optimized purely for short-term utility, selects at each step the action that maximizes immediate engagement. In doing so, it tends to favor the Group~$2$ that shows higher current activity. This induces a feedback loop: the advantaged Group~$2$, receiving more exposure, further increases its engagement, while the disadvantaged Group~$1$ receives less exposure and becomes progressively less active, see Figure~\ref{fig:representation_bias}. As highlighted in \cite{NP-JB-EE-GDP-SB-AH:23,Liu_delayedImpact_ICML18}, such dynamics can lead to persistent disparities or even long-term disengagement of the disadvantaged group.


In contrast, a fairness-aware recommender treats fairness as a \emph{long-term control objective}. 
In the previous section, we operated within a fixed dynamical regime to meet exposure targets.
Here, we explicitly account for how current decisions reshape future system states.
This leads to a sequential optimization problem of the form
\[
\max_{\beta} \; \sum_{t=0}^{T} \gamma^t \, \mathcal{U}(v_t, \beta(v_t))
\quad \text{s.t.} \quad v_{t+1} = f(v_t, \beta(v_t)),
\]
where $\beta$ is the recommendation policy, $\mathcal{U}$ encodes both utility and fairness considerations, and $f$ describes how recommendations influence future engagement.
Under such a policy, the system may intentionally allocate more exposure to the disadvantaged Group~$1$, even at the cost of short-term utility, in order to improve its future engagement and avoid long-term inequality. As a result, the system can converge to a qualitatively different steady state, where engagement is sustained across both groups. This illustrates the key insight of \cite{Liu_delayedImpact_ICML18}: optimizing for long-term objectives can fundamentally change the fairness--utility trade-off, emphasizing the importance of incorporating dynamics into the design of fair RS. These approaches show that sacrificing short-term optimality can steer the system toward more equitable steady states, effectively reshaping the underlying population dynamics \cite{Liu_delayedImpact_ICML18,zhang2020fairdecisions,hu2018shortterm,zhang2020longterm}.


The formulation above also connects naturally to the fast-timescale framework discussed earlier: the policy $\beta$ can represent, for instance, an exposure target. The fast controller would handle individual ranking decisions, while the long-term design problem focuses on selecting the best exposure target or fairness constraint to impose.
Importantly, achieving such outcomes requires adaptive and dynamic notions of fairness. Static constraints  are generally insufficient, as they ignore the evolving nature of item attributes, user preferences, and group composition \cite{WY-MW-ZM-LY-MS:23,biases_review,fairness_dynamic_ML}. In contrast, dynamic fairness approaches, such as those based on constrained sequential decision-making or control, continuously adjust the policy to maintain fairness as the environment evolves, enabling both efficiency and equity in the long run \cite{morik2020controlling,learning_rs,SC-GDP-GB-FD:25,fairness_socialequality,fairness_equilibria}.

Across diverse settings like hiring, lending, and education the literature has shown that the impact of fairness constraints depends critically on the underlying dynamics: they can help, harm, or leave unaffected the groups they aim to protect \cite{fairness_socialequality, zhang2020longterm}. 
For instance, fairness constraints can sometimes break natural equality or reduce incentives for disadvantaged groups to improve their qualifications, while in other regimes they improve long-run social welfare. These nuanced outcomes underscore that there is no universally beneficial fairness intervention; the right choice depends on the feedback structure of the system at hand.
Overall, the literature suggests that fairness interventions, when properly designed to account for temporal feedback and system dynamics, can do more than mitigate harm: they can actively guide platforms toward socially desirable equilibria. This highlights a fundamental shift from viewing fairness as a static constraint to understanding it as a long-term system design principle\cite{fairness_dynamic_ML}, where the goal is not only fair decisions at each step, but fair outcomes in the limit \cite{clickbait,rs_regret,rs_improvingquality}.

\section{CONCLUSION}
We reframed RS through the lens of control theory. By interpreting recommendation pipelines as dynamical systems, we demonstrated that fairness interventions, when embedded directly into the system’s objective, can stabilize user–creator interactions, mitigate runaway feedback loops, and ultimately improve performance over time. On the user side, incorporating the social network proved essential: fairness cannot be meaningfully assessed without acknowledging the social influence users receive within their network. On the creator side, we showed that fairness-aware objectives can prevent the concentration dynamics that otherwise degrade diversity and opportunities.
A key insight emerging from the dynamical perspective is that fairness is not necessarily a fixed quantity to be traded off against accuracy. Instead, the “right” fairness definition depends on the specific behavioral dynamics of users and creators. Even under identical system goals, different environments call for different fairness formulations. This underscores the need for fairness research that is sensitive to temporal evolution, feedback, and adaptation dimensions that static metrics cannot capture \cite{fairness_dynamic_ML}.
The simulations in this paper are intentionally simple. This should not be taken to imply that real‑world systems are simple; rather, our aim is to show that even under highly simplified dynamics, the feedback‑driven processes underlying recommender systems can already exhibit the instabilities and amplification phenomena that motivate deeper control‑theoretic analysis.
Our findings call for deeper engagement from the control community. RS, and automated decision-making systems more broadly, are feedback-driven processes. They exhibit the instability, amplification, and fragility that control theory was built to understand and to counter. Yet the field has largely been shaped by HCI and machine learning traditions, where models are often empirical, data-driven, and not designed to capture the underlying system dynamics. As a result, today’s RS operate with models far less reliable than those in classical control domains, making them more fragile and more susceptible to harmful emergent behavior.

This is precisely why the moment is ripe for control theory to step in. Machine learning researchers increasingly acknowledge that RS are control problems\cite{fairness_dynamic_ML}. The next era of RS will be defined by our ability to model, analyze, and regulate them as such. Doing so requires confronting major challenges: the scarcity of validated dynamical models, the difficulty of obtaining longitudinal and counterfactual data, and the need for fairness definitions that adapt to evolving system behavior rather than freezing it in time. 
But these challenges also mark the frontier. If the control community embraces them, we can reshape the foundations of RS, transforming fairness from a reactive patch into a principled, system-level design goal. The opportunity is not merely to improve algorithms, but to redefine how large-scale digital ecosystems are governed. In doing so, we can help build systems that are not only more equitable, but also more stable, more resilient, and ultimately more aligned with the long-term well-being of both users and creators.

\section{ACKNOWLEDGMENT}

S.D. was partly supported by NSF CCF 2312774, NSF OAC-2311521, NSF IIS-2442137, a gift to the LinkedIn-Cornell Bowers CIS Strategic Partnership, and an AI2050 Early Career Fellowship program at Schmidt Sciences. 
P.F.~was partly supported by the French National Research Agency through project FeedingBias (ANR-22-CE380017-01).

\section{Author Information}

\begin{IEEEbiography}{{G}iulia De Pasquale}{\,}(g.de.pasquale@tue.nl) received
the B.Sc. degree in information engineering, the M.Sc. degree in control engineering, and the Ph.D. degree in control systems from the University of Padova, Padua, Italy, in 2017, 2019, and 2023, respectively. She is currently an Assistant Professor with the Control Systems Group, Einhoven University of Technology, Eindhoven, The Netherlands. In 2022, she was a Visiting Research Scholar with the University of California Santa Barbara, Santa Barbara, CA, USA. From 2022 to 2024, she was a Postdoctoral Researcher with the Automatic Control Laboratory, ETH Zürich, Zürich, Switzerland. Her current research interests include modeling, analysis, optimization, and control of networked sociotechnical systems. She is a Member of IEEE. 
\end{IEEEbiography}
\begin{IEEEbiography}{{S}arah Dean}{\,}(sdean@cornell.edu) is an assistant professor in the Department of Computer Science at Cornell University, NY, USA. She studies the interplay between  optimization, machine learning, and dynamics in real-world systems. Her research focuses on  understanding the fundamentals of data-driven methods for control and decision-making, inspired  by applications ranging from robotics to recommendation systems. She completed her postdoctoral  research at the University of Washington and earned her M.S. and Ph.D. in electrical engineering and computer science  at the University of California, Berkeley. Dean received her B.S.E. in electrical engineering and mathematics from the  University of Pennsylvania.
\end{IEEEbiography}

\begin{IEEEbiography}{{P}aolo Frasca}{\,}(paolo.frasca@gipsa-lab.fr) is a CNRS researcher affiliated with GIPSA-Lab (Grenoble, France), where since 2021 he has been leading the DANCE research team devoted to Dynamics and Control of Networks.
His research interests cover the theory of networks and of control and learning systems, with main applications in transportation systems, social networks, and more generally socio-technical systems. 
He received the PhD degree from Politecnico di Torino (Turin, Italy), in 2009.
\end{IEEEbiography}

\bibliography{references}

\endarticle

\end{document}